\documentclass[twocolumn]{aastex7}
\usepackage{xspace,amsmath,amssymb,multirow}

\newcommand{\Lya}{Ly\ensuremath{\alpha}\xspace}
\newcommand{\oiii}{[O\,\textsc{iii}]}
\newcommand{\Feii}{Fe\,\textsc{ii}}

\newcommand{\src}{CAPERS-LRD-z9}

\shorttitle{}
\shortauthors{}

\begin{document}

\title{\src{}: A Gas Enshrouded Little Red Dot Hosting a Broad-line AGN at $\boldsymbol{z=9.288}$}

\correspondingauthor{Anthony~J.~Taylor}
\email{anthony.taylor@austin.utexas.edu}

\author[0000-0003-1282-7454]{Anthony J. Taylor}
\email{anthony.taylor@austin.utexas.edu}
\affiliation{Department of Astronomy, The University of Texas at Austin, Austin, TX, USA}

\author[0000-0002-5588-9156]{Vasily Kokorev}
\email{vkokorev@utexas.edu}
\affiliation{Department of Astronomy, The University of Texas at Austin, Austin, TX, USA}

\author[0000-0002-8360-3880]{Dale D. Kocevski}
\email{dkocevsk@colby.edu}
\affiliation{Department of Physics and Astronomy, Colby College, Waterville, ME 04901, USA}

\author[0000-0003-3596-8794]{Hollis B. Akins}
\email{hollis.akins@utexas.edu}
\affiliation{Department of Astronomy, The University of Texas at Austin, Austin, TX, USA}

\author[0000-0002-3736-476X]{Fergus Cullen}
\email{fergus.cullen@ed.ac.uk}
\affiliation{Institute for Astronomy, University of Edinburgh, Royal Observatory, Edinburgh EH9 3HJ, UK}

\author[0000-0001-5414-5131]{Mark Dickinson}
\email{mark.dickinson@noirlab.edu}
\affiliation{NSF's National Optical-Infrared Astronomy Research Laboratory, 950 N. Cherry Ave., Tucson, AZ 85719, USA}

\author[0000-0001-8519-1130]{Steven L. Finkelstein}
\email{stevenf@astro.as.utexas.edu}
\affiliation{Department of Astronomy, The University of Texas at Austin, Austin, TX, USA}

\author[0000-0002-7959-8783]{Pablo Arrabal Haro}
\email{pablo.arrabalharo@nasa.gov}
\altaffiliation{NASA Postdoctoral Fellow}
\affiliation{Astrophysics Science Division, NASA Goddard Space Flight Center, 8800 Greenbelt Rd, Greenbelt, MD 20771, USA}

\author[0000-0003-0212-2979]{Volker Bromm}
\email{bromm@astro.as.utexas.edu}
\affiliation{Department of Astronomy, The University of Texas at Austin, Austin, TX, USA}

\author[0000-0002-7831-8751]{Mauro Giavalisco}
\email{mauro@umass.edu}
\affiliation{University of Massachusetts Amherst, 710 North Pleasant Street, Amherst, MA 01003-9305, USA}

\author[0000-0001-9840-4959]{Kohei Inayoshi}
\email{inayoshi0328@gmail.com}
\affiliation{Kavli Institute for Astronomy and Astrophysics, Peking University, Beijing 100871, China}

\author[0000-0002-0000-2394]{St{\'e}phanie Juneau}
\email{stephanie.juneau@noirlab.edu}
\affiliation{NSF's National Optical-Infrared Astronomy Research Laboratory, 950 N. Cherry Ave., Tucson, AZ 85719, USA}

\author[0000-0002-9393-6507]{Gene C. K. Leung}
\email{gckleung@mit.ude}
\affiliation{MIT Kavli Institute for Astrophysics and Space Research, 77 Massachusetts Ave., Cambridge, MA 02139, USA}

\author[0000-0003-4528-5639]{Pablo G. P\'erez-Gonz\'alez}
\email{pgperez@cab.inta-csic.es}
\affiliation{Centro de Astrobiolog\'{\i}a (CAB), CSIC-INTA, Ctra. de Ajalvir km 4, Torrej\'on de Ardoz, E-28850, Madrid, Spain}

\author[0000-0002-6748-6821]{Rachel S. Somerville}
\email{rsomerville@flatironinstitute.org}
\affiliation{Center for Computational Astrophysics, Flatiron Institute, 162 5th Avenue, New York, NY, 10010, USA}

\author[0000-0002-1410-0470]{Jonathan R. Trump}
\email{jonathan.trump@uconn.edu}
\affiliation{Department of Physics, 196 Auditorium Road, Unit 3046, University of Connecticut, Storrs, CT 06269, USA}

\author[0000-0001-5758-1000]{Ricardo O. Amor\'{i}n}
\email{amorin@iaa.csic.es}
\affiliation{Instituto de Astrof\'{i}sica de Andaluc\'{i}a (CSIC), Apartado 3004, 18080 Granada, Spain}

\author[0000-0001-6813-875X]{Guillermo Barro}
\email{gbarro@pacific.edu}
\affiliation{Department of Physics, University of the Pacific, Stockton, CA 90340 USA}

\author[0000-0002-4193-2539]{Denis Burgarella}
\email{denis.burgarella@lam.fr}
\affiliation{Aix Marseille Univ, CNRS, CNES, LAM Marseille, France}

\author[0000-0001-5384-3616]{Madisyn Brooks}
\email{madisyn.brooks@uconn.edu}
\affiliation{Department of Physics, 196 Auditorium Road, Unit 3046, University of Connecticut, Storrs, CT 06269, USA}

\author[0000-0000-0000-0000]{Adam C. Carnall}
\email{adamc@roe.ac.uk}
\affiliation{Institute for Astronomy, University of Edinburgh, Royal Observatory, Edinburgh EH9 3HJ, UK}

\author[0000-0002-0930-6466]{Caitlin M. Casey}
\email{cmcasey@ucsb.edu}
\affiliation{Department of Physics, University of California, Santa Barbara, CA 93106, USA}
\affiliation{Cosmic Dawn Center (DAWN), Niels Bohr Institute, University of Copenhagen, Jagtvej 128, K{\o}benhavn N, DK-2200, Denmark}

\author[0000-0001-8551-071X]{Yingjie Cheng}
\email{yingjiecheng@umass.edu}
\affiliation{University of Massachusetts Amherst, 710 North Pleasant Street, Amherst, MA 01003-9305, USA}

\author[0000-0002-0302-2577]{John Chisholm}
\email{chisholm@austin.utexas.edu}
\affiliation{Department of Astronomy, The University of Texas at Austin, Austin, TX, USA}

\author[0000-0003-4922-0613]{Katherine Chworowsky}
\email{k.chworowsky@utexas.edu}
\altaffiliation{NSF Graduate Fellow}
\affiliation{Department of Astronomy, The University of Texas at Austin, Austin, TX, USA}

\author[0000-0001-8047-8351]{Kelcey Davis}
\email{kelcey.davis@uconn.edu}
\altaffiliation{NSF Graduate Research Fellow}
\affiliation{Department of Physics, 196 Auditorium Road, Unit 3046, University of Connecticut, Storrs, CT 06269, USA}

\author[0000-0002-7622-0208]{Callum T. Donnan}
\email{callum.donnan@noirlab.edu}
\affiliation{NSF's National Optical-Infrared Astronomy Research Laboratory, 950 N. Cherry Ave., Tucson, AZ 85719, USA}

\author[0000-0002-1404-5950]{James S. Dunlop}
\email{James.Dunlop@ed.ac.uk}
\affiliation{Institute for Astronomy, University of Edinburgh, Royal Observatory, Edinburgh EH9 3HJ, UK}

\author[0000-0001-7782-7071]{Richard S. Ellis}
\email{ richard.ellis@ucl.ac.uk}
\affiliation{Department of Physics and Astronomy, University College London, Gower Street, London WC1E 6BT, UK}

\author[0000-0003-0531-5450]{Vital Fern\'andez}
\email{vgf@umich.edu}
\affiliation{Michigan Institute for Data Science, Unversity of Michigan, 500 Church Street, Ann Arbor, MI 48109, USA}

\author[0000-0001-7201-5066]{Seiji Fujimoto}
\email{fujimoto@utexas.edu}
\affiliation{Department of Astronomy, The University of Texas at Austin, Austin, TX, USA}

\author[0000-0001-9440-8872]{Norman A. Grogin}
\email{nagrogin@stsci.edu}
\affiliation{Space Telescope Science Institute, 3700 San Martin Drive, Baltimore, MD 21218, USA}

\author[0000-0003-4242-8606]{Ansh R. Gupta}
\email{anshrg@utexas.edu}
\altaffiliation{NSF Graduate Research Fellow}
\affiliation{Department of Astronomy, The University of Texas at Austin, Austin, TX, USA}

\author[0000-0001-6145-5090]{Nimish P. Hathi}
\email{nhathi@stsci.edu}
\affiliation{Space Telescope Science Institute, 3700 San Martin Drive, Baltimore, MD 21218, USA}

\author[0000-0003-1187-4240]{Intae Jung}
\email{ijung@stsci.edu}
\affiliation{Space Telescope Science Institute, 3700 San Martin Drive, Baltimore, MD 21218, USA}

\author[0000-0002-3301-3321]{Michaela Hirschmann}
\email{michaela.hirschmann@epfl.ch}
\affiliation{Institute of Physics, Laboratory of Galaxy Evolution, Ecole Polytechnique Federale de Lausanne (EPFL), Observatoire de Sauverny, 1290 Versoix, Switzerland}

\author[0000-0001-9187-3605]{Jeyhan S. Kartaltepe}
\email{jeyhan@astro.rit.edu}
\affiliation{Laboratory for Multiwavelength Astrophysics, School of Physics and Astronomy, Rochester Institute of Technology, 84 Lomb Memorial Drive, Rochester, NY 14623, USA}

\author[0000-0002-6610-2048]{Anton M. Koekemoer}
\email{koekemoer@stsci.edu}
\affiliation{Space Telescope Science Institute, 3700 San Martin Drive, Baltimore, MD 21218, USA}

\author[0000-0003-2366-8858]{Rebecca L. Larson}
\email{rlarson@stsci.edu}
\affiliation{Space Telescope Science Institute, 3700 San Martin Drive, Baltimore, MD 21218, USA}

\author[0000-0003-0486-5178]{Ho-Hin Leung}
\email{hleung2@roe.ac.uk}
\affiliation{Institute for Astronomy, University of Edinburgh, Royal Observatory, Edinburgh EH9 3HJ, UK}

\author[0000-0003-1354-4296]{Mario Llerena}
\email{mario.llerenaona@inaf.it}
\affiliation{INAF - Osservatorio Astronomico di Roma, via Frascati 33, 00078, Monte Porzio Catone, Italy}

\author[0000-0003-1581-7825]{Ray A. Lucas}
\email{lucas@stsci.edu}
\affiliation{Space Telescope Science Institute, 3700 San Martin Drive, Baltimore, MD 21218, USA}

\author[0000-0003-4368-3326]{Derek J. McLeod}
\email{derek.mcleod@ed.ac.uk}
\affiliation{Institute for Astronomy, University of Edinburgh, Royal Observatory, Edinburgh EH9 3HJ, UK}

\author[0000-0000-0000-0000]{Ross McLure}
\email{rmclure@ed.ac.uk}
\affiliation{Institute for Astronomy, University of Edinburgh, Royal Observatory, Edinburgh EH9 3HJ, UK}

\author[0000-0002-8951-4408]{Lorenzo Napolitano}
\email{lorenzo.napolitano@inaf.it}
\affiliation{INAF - Osservatorio Astronomico di Roma, via Frascati 33, 00078, Monte Porzio Catone, Italy}
\affiliation{Dipartimento di Fisica, Università di Roma Sapienza, Città Universitaria di Roma - Sapienza, Piazzale Aldo Moro, 2, 00185, Roma, Italy}

\author[0000-0001-7503-8482]{Casey Papovich}
\email{papovich@tamu.edu}
\affiliation{Department of Physics and Astronomy, Texas A\&M University, College Station, TX, 77843-4242 USA}
\affiliation{George P.\ and Cynthia Woods Mitchell Institute for Fundamental Physics and Astronomy, Texas A\&M University, College Station, TX, 77843-4242 USA}

\author[0000-0002-0827-9769]{Thomas M. Stanton}
\email{t.stanton@ed.ac.uk}
\affiliation{Institute for Astronomy, University of Edinburgh, Royal Observatory, Edinburgh EH9 3HJ, UK}

\author[0000-0002-9909-3491]{Roberta Tripodi}
\email{roberta.tripodi@inaf.it}
\affiliation{INAF - Osservatorio Astronomico di Roma, via Frascati 33, 00078, Monte Porzio Catone, Italy}
\affiliation{IFPU - Institute for Fundamental Physics of the Universe, Via Beirut 2, I-34014 Trieste, Italy}

\author[0000-0002-9373-3865]{Xin Wang}
\email{xwang@ucas.ac.cn}
\affiliation{School of Astronomy and Space Science, University of Chinese Academy of Sciences (UCAS), Beijing 100049, China}
\affiliation{National Astronomical Observatories, Chinese Academy of Sciences, Beijing 100101, China}
\affiliation{Institute for Frontiers in Astronomy and Astrophysics, Beijing Normal University, Beijing 102206, China}

\author[0000-0003-3903-6935]{Stephen M.~Wilkins}
\email{s.wilkins@sussex.ac.uk}
\affiliation{Astronomy Centre, University of Sussex, Falmer, Brighton BN1 9QH, UK}
\affiliation{Institute of Space Sciences and Astronomy, University of Malta, Msida MSD 2080, Malta}

\author[0000-0003-3466-035X]{{L. Y. Aaron} {Yung}}
\email{yung@stsci.edu}
\affiliation{Space Telescope Science Institute, 3700 San Martin Drive, Baltimore, MD 21218, USA}

\author[0000-0002-7051-1100]{Jorge A. Zavala}
\email{jzavala@umass.edu}
\affiliation{University of Massachusetts Amherst, 710 North Pleasant Street, Amherst, MA 01003-9305, USA}

\begin{abstract}
We present \src{}, a little red dot (LRD) which we confirm to be a $z=9.288$ broad-line AGN (BLAGN). First identified as a high-redshift LRD candidate from PRIMER NIRCam photometry, follow-up NIRSpec/PRISM spectroscopy of \src{} from the CANDELS-Area Prism Epoch of Reionization Survey (CAPERS) has revealed a broad $3500$~km~s$^{-1}$ full-width-half-maximum H$\beta$ emission line and narrow \oiii$\lambda\lambda4959,5007$ lines, indicative of a BLAGN. Based on the broad H$\beta$ line, we compute a canonical black-hole mass of $\log(M_{\textrm{BH}}/M_{\odot})=7.58\pm0.15$, although full consideration of systematic uncertainties yields a conservative range of $6.65<\log(M_{\textrm{BH}}/M_{\odot})<8.50$. These observations suggest that either a massive black hole seed, or a lighter stellar remnant seed undergoing periods of super-Eddington accretion, is necessary to grow such a massive black hole in $\lesssim500$~Myr of cosmic time. \src{} exhibits a strong Balmer break, consistent with a central AGN surrounded by dense ($\sim 10^{10}\textrm{~cm}^{-3}$) neutral gas. We model \src{} using \texttt{Cloudy} to fit the emission red-ward of the Balmer break with a dense gas-enshrouded AGN, and \texttt{bagpipes} to fit the rest-ultraviolet emission as a host-galaxy stellar population. This upper limit on the stellar mass of the host galaxy ($<~10^9\,{\rm M_\odot}$) implies that the black-hole to stellar mass ratio may be extremely large, possibly $>5\%$ (although systematic uncertainties on the black-hole mass prevent strong conclusions).  However, the shape of the UV continuum differs from typical high-redshift star-forming galaxies, indicating that this UV emission may also be of AGN origin, and hence the true stellar mass of the host may be still lower. 
\end{abstract} 

\section{Introduction}\label{sec:intro}

In the years following its launch, \textit{JWST} has uncovered previously inaccessible populations of galaxies. These objects include an unprecedented population of galaxies at redshifts of $z>10$ \citep[e.g.,][]{finkelstein22maisie,castellano22,naidu22b,arrabalharo23,robertson23,wang23,carniani24,donnan2024,kokorev25,Zavala25}, and a surprisingly large population of broad-line active galactic nuclei (BLAGN) at $z\gtrsim6$ \citep[e.g.,][]{larson23,bunker23,kocevski23,harikane23, kocevski25,Maiolino_2023,akins24,matthee24,furtak24,juodzbalis24,kokorev23,tripodi24,taylor24,wang24a,juodzbalis24,juodzbalis25,akins25}. 

One of the more enigmatic populations identified by \textit{JWST} are the red compact objects that have been named ``Little Red Dots'' (LRDs; \citealt{matthee24}).  These sources were first identified photometrically due to their compact morphology and unique ``v-shaped'' spectral energy distributions (SEDs), which feature a steep red continuum in the rest-frame optical, with relatively blue colors in the rest-frame UV \citep{kocevski23,Barro23,labbe23}.  Subsequent spectroscopic observations have revealed that $\gtrsim60\%$ of these sources exhibit broad Balmer emission lines, which when coupled with observed narrow forbidden lines is indicative of BLAGN \citep{kocevski23,kokorev23,furtak24,greene24,matthee24,kocevski25}. While alternative explanations for these broad-line features have been proposed \citep[e.g.,][]{Baggen24}, several lines of evidence such as high-ionization lines \citep{labbe24, tripodi24, akins25} and Balmer absorption \citep{matthee24,ji25} point to an AGN powering a significant fraction of the rest-frame optical emission.

LRDs have been shown to be numerous, making up $\sim20-30\%$ of the BLAGN identified by JWST at $z>3$ \citep{harikane23,greene24,maiolino24,taylor24}. Their number density matches or exceeds the total pre-\textit{JWST} expected contribution of UV-faint ($M_{UV}>-21$) AGN to the UV luminosity function at $4.5<z<8.5$, and LRDs represent $\sim1\%$ of the total galaxy UVLF over the same range \citep{finkelsteinbagley22,kokorev24,kocevski25,taylor24}. Under the broad-line AGN interpretation, LRDs host supermassive black holes (SMBHs) with some reaching $M_{\textrm{BH}}> 10^7 M_{\odot}$ within the first Gyr of cosmic time. These objects thus provide insight into black hole seeding and growth in the early universe \citep[e.g.,][]{Smith2019,Woods2019,Inayoshi2020,Regan2024,Jeon_SMBH2025}. 

However, analysis of the rest-frame near-infrared (near-IR), or observed mid-infrared (mid-IR), emission of LRDs further complicates their physical interpretation. Observations with JWST/MIRI have shown a surprisingly flat continuum at rest-frame $1-3 \mu$m \citep[e.g.,][]{Williams24, PerezGonzalez24, akins24, leung24, barro24b}, suggesting a lack of a hot dusty torus commonly seen in reddened AGNs at up to $z \sim 6$ \citep[e.g.,][]{barvainis87, lyu22}. Full SED analysis using NIRCam and MIRI shows either no correlation \citep{leung24} or a negative correlation \citep{barro24b} between the apparent extinction in LRDs and their rest-MIR emission, indicating an alternative source of obscuration than hot dust. Moreover, applying a conventional AGN SED model with dust obscuration implies highly over-massive black holes in LRDs, hinting at either some stellar contribution or super-Eddington accretion \citep{leung24}.

A subset of LRDs with deep  \textit{JWST}/NIRSpec PRISM spectroscopy exhibit strong Balmer breaks \citep[e.g.,][]{wang24, setton24}. While this initially suggested that the emission at these wavelengths must be dominated by stars, some of these objects exhibit greater break strengths than are possible from an evolved stellar population alone \citep[e.g.,][]{Williams24,labbe24,ji25,degraaff25,naidu25}, thus requiring an alternative explanation. \cite{inayoshi25} suggest that generating such breaks is in fact possible without stellar emission, by instead modeling an AGN enshrouded by a cocoon of dense neutral gas. In this model, the high column density of this gas results in a significant population of $n=2$ collisionally excited state hydrogen atoms that resonantly scatter Balmer line emission, and absorb and reprocess light at wavelengths less than the Balmer limit ($\sim$3650~\AA{}), much like photons blueward of the Lyman limit in less-dense gas where the bulk of hydrogen atoms are in the ground state. This results in a stronger Balmer break, and narrow absorption features superimposed upon the broad Balmer emission lines. These absorption features have indeed been identified in several LRDs \citep{matthee24,kocevski25,taylor24,wang24a,ji25,degraaff25,rusakov25,naidu25}. This theory also helps explain the lack of X-ray emission exhibited by LRDs due to Compton-thick absorption \citep{kocevski25,yue24xray,maiolino25} as well as the higher Balmer decrements observed from broad lines compared to narrow lines \citep[e.g.,][]{brooks24,naidu25}. 

The origin and growth histories of these supermassive black holes remains an open puzzle \citep{Inayoshi2020,Volonteri2021}. The most commonly considered scenarios for creating seed black holes can be divided into two categories: 1) Light seeds ($m_{\rm seed} \sim 100\,M_{\odot}$), left behind as stellar remnants of massive Population~III (Pop~III) stars, and 2) Heavy seeds, which may form via runaway collisions in dense environments such as stellar clusters (thought to create seeds with $m_{\rm seed} \sim 10^3$--$10^4\, M_{\odot}$, which may rapidly grow to $\sim 10^5$--$10^6\,M_{\odot}$ via mergers), or via direct collapse of primordial gas clouds under special conditions such as strong Lyman-Werner (LW) radiation fields, high baryon-DM streaming velocity, or gas-rich mergers, leading to seeds with masses $m_{\rm seed} \sim 10^5$--$10^6\,M_{\odot}$ \citep[see review by][and references therein]{Inayoshi2020}. A more exotic origin for seed BHs, involving a qualitatively different evolutionary sequence, is primordial black holes that form shortly after the Big Bang \citep[e.g.,][]{LiuBromm2022,Zhang_PBH2025,Ziparo_PBH2025,Matteri2025}. It is possible, even likely, that multiple mechanisms contribute to seeding the black hole population that we observe. 

In order to reach the masses of the detected high redshift black holes, light seeds must grow at or above the Eddington rate with a high duty cycle from an early epoch ($z\gtrsim 20$). Heavier seeds can grow at more moderate rates. For the broader category of non-LRD broad line high redshift AGN, the estimated BH masses appear to make many of these objects ``over-massive'' compared to their stellar hosts, with black hole to stellar mass ratios far above the typical $M_{\textrm{BH}}/M_*\sim0.1$\% exhibited by local dormant black holes and AGN \citep{agarwal13,pacucci23}. However, we note that analyzing this ``over-massiveness'' on a population level may be subject to Lauer bias \citep[the tendency to observe a greater number of massive black holes as outliers in intrinsically common moderate mass galaxies, rather than in intrinsically rarer high-mass galaxies;][]{lauer07,lisilverman25}. The stellar masses of LRDs are particularly uncertain, given that it is unknown whether the observed UV light comes from the AGN or the stars, but some estimates suggest that they may have some of the most extreme over-massive BHs, with some approaching $\sim10-100$\% of the host galaxy's stellar mass \citep[e.g.][]{Maiolino_2023,kokorev23,kocevski25,Chen_2025}. Heavy seeds would provide one obvious explanation for over-massive black holes, but there are other explanations such as periods of more rapid growth of the BH relative to the stars, or selection effects. Pushing black hole detections to earlier cosmic epochs, closer to the presumed epoch of BH seeding, is one of the most promising avenues to disentangle the degeneracies between seeding and accretion physics. 

These numerous open questions have prompted ongoing campaigns for spectroscopic followup of photometrically identified LRDs, especially those at high photometric redshifts $(z\gtrsim9)$. Here, we present \src{}, the highest redshift spectroscopically confirmed LRD and BLAGN yet discovered. The source was first identified as a high-$z$ LRD candidate in \cite{kocevski25} and subsequently also selected by \cite{barro24b} and \cite{akins24}.  It was also independently selected as a galaxy candidate at $z = 9-10$ by \citet{donnan2024}.  We now observe this source using NIRSpec observations taken as part of the CANDELS-Area Prism Epoch of Reionization Survey (CAPERS) program (GO-6368; PI. M. Dickinson).

This paper is organized as follows. First, in \S\ref{sec:observations}, we describe the \textit{JWST} NIRCam, NIRSpec, and MIRI data used in the discovery and follow-up of \src{}. In \S\ref{sec:analysis}, we describe our measurements of the spectroscopic redshift, emission-line properties, morphology, Balmer decrement, and Balmer break in \src{}. In \S\ref{sec:results}, we compute the black hole and stellar masses of \src{}, and we discuss the implications of these properties and the applicability of the dense neutral gas model of LRDs to this source. Finally, we summarize our work in \S\ref{sec:summary}. We assume a flat $\Lambda$ cold dark matter cosmology with $\Omega_m$=$0.3$, $\Omega_\Lambda$=$0.7$, and $H_0$=$70$~km~s$^{-1}$~Mpc$^{-1}$ throughout. All magnitudes are given in the AB magnitude system \citep{oke83}, where an AB magnitude is defined by $m_{\rm AB}$=$-2.5\log f_\nu - 48.60$. Here, $f_\nu$ is the specific flux density of the source in units of erg~cm$^{-2}$~s$^{-1}$~Hz$^{-1}$.

\section{Observations and Data Reduction}\label{sec:observations}

\src{} was first identified as a potential very high redshift LRD from the {\it JWST} NIRCam imaging provided by the PRIMER survey \citep{dunlop_2021_primer} in the COSMOS field \citep{kocevski25,barro24b,akins24}. The LRD selection was based on an analysis of the rest-frame UV and optical spectral energy distribution (SED) slopes following determination of the source's photometric redshift with \texttt{EAZY} \citep{brammer08}. As potentially the highest redshift LRD discovered to date, \src{} was identified as a high-value target for CAPERS, and therefore it was selected to receive a slitlet and a full-depth CAPERS integration (see \S\ref{sec:nirspec}). While the full details of CAPERS target selection, data reduction, and program design will be described in future works, we summarize the observations below. 

\subsection{NIRSpec Observations}\label{sec:nirspec}

CAPERS is a {\it JWST} Cycle-3 Legacy program which uses NIRSpec/MSA/PRISM to observe very high-redshift galaxies, AGN candidates, and other objects of community interest in three of the CANDELS legacy fields \citep{grogin11, koekemoer11} as identified from the deep multi-band NIRCam imaging provided by the Public Release IMaging for Extragalactic Research (PRIMER; Dunlop et al.\ in preparation) and the Cosmic Evolution Early Release Science \citep[CEERS;][]{finkelstein25} {\it JWST} surveys. 

CAPERS is targeting seven MSA pointings in each of the PRIMER-UDS, PRIMER-COSMOS, and Extended Groth Strip (EGS) fields, observing three MSA configurations at each pointing. This multi-configuration approach allows CAPERS to observe high-value targets in multiple configurations to increase their exposure times, while simultaneously observing a large number of other targets in each individual configuration to increase the overall sample yield. As a high-value target, \src{} (a.k.a.\ CAPERS-COSMOS-119334) was observed in all three configurations executed in CAPERS COSMOS pointing P4. These observations targeting \src{} were executed on 2025 April 15 using two iterations of a standard three-shutter nodding scheme per MSA configuration for a total of 18 exposures at a NIRSpec aperture position angle (east of north) of 246.585$^{\circ}$. Each of these 18 nods was observed for a single 13 group integration using the \texttt{NRSIRS2} readout pattern for a combined effective exposure time of 17069~s (4.74~hrs). 

We reduced the spectroscopic data using the \textit{JWST} Calibration Pipeline\footnote{\url{https://github.com/spacetelescope/jwst}} \citep{bushouse23} version 1.17.1 and CRDS version \texttt{1350.pmap}. We largely used the default configuration for the pipeline with a few notable exceptions. First, we enabled the \texttt{clean\_flicker\_noise} step (using the ``median" method\footnote{Based on image1overf.py \url{https://github.com/chriswillott/jwst}}) in the \texttt{calwebb\_detector1} stage to remove the effects of 1/f noise in the count rate maps. We also employed a modified flat-field file in the \texttt{calwebb\_spec2} stage \citep[see][for details]{arrabalharo23}. Beyond these changes, we ran the \texttt{calwebb\_spec3} stage using the CRDS defaults to produce a final 2D spectrum (see Figure~\ref{fig:fullspectrum}, middle panel). 

We next used a custom optimal extraction \citep{horne86} to best detect and extract the signal from the source in the 2D spectrum to produce a 1D spectrum. Finally, we calibrated this 1D spectrum to the \textit{JWST}/NIRCam photometry provided by the PRIMER survey (\S\ref{sec:nircam}) by computing the total flux in the spectrum within each of the F150W, F200W, F277W, F356W, and F444W NIRCam passbands. Specifically, we compared these measured fluxes to the photometric measurements of the object from NIRCam and fitted a third-order Chebyshev polynomial to the ratios between the spectrum and the photometry (as in \texttt{BAGPIPES}; \citealt{carnall18}). This correction peaks at a factor of $\sim$0.9 in F150W and quickly flattens to $\sim$0.7 from F200W through F444W. We then multiplied the 1D spectrum by the resulting fitted correction curve to produce our final calibrated 1D spectrum (see Figure~\ref{fig:fullspectrum}, lower panel). This calibration is necessary to correct for the effects of NIRSpec path-loss and automated corrections applied in the \textit{JWST} Calibration Pipeline that are not tied to NIRCam data.

\subsection{NIRCam Photometry}\label{sec:nircam}

The analysis in this paper makes use of \textit{JWST}/NIRCam photometry in the PRIMER-COSMOS field. This imaging is an internal reduction from the PRIMER team (internal version 1.0). The PRIMER imaging data were reduced using the PRIMER Enhanced NIRCam Image Processing Library (PENCIL; Magee et al., in preparation, Dunlop et al., in preparation) software. 

The photometry was measured via a similar process as in \cite{finkelstein24}.  Briefly, for each band of imaging, this methodology PSF-matches images (using empirical PSFs) with smaller PSFs than F277W to that band, and derives correction factors for images with larger PSFs (via PSF-matching F277W to a given larger PSF).  Kron apertures are used to measure colors to optimize signal-to-noise for high-redshift galaxies.  Total fluxes are estimated by deriving an aperture correction in the F277W band as the ratio between the flux in the larger Kron aperture and the custom smaller aperture, with a residual aperture correction (typically $<$10\%) derived via source-injection simulations. We show cutouts of the imaging of \src{} and list the measured magnitudes in Figure~\ref{fig:fullspectrum} (upper panel).

\subsection{MIRI Photometry}\label{sec:miri}
We searched the Barbara~A.~Mikulski Archive for Space Telescopes (MAST) for MIRI data at the position of \src{}. It was located in a gap of the PRIMER observations but was covered by the COSMOS-3D survey (PID 5893) on 2025 April 20. Raw data were downloaded from MAST and reduced with the Rainbow MIRI software \citep{PerezGonzalez24} using the \textit{JWST} Calibration Pipeline version v1.18.0, reference files in \texttt{jwst\_1364.pmap}. The Rainbow software implements a super-background strategy to remove and homogenize the background, improving the detectability of faint sources. We refer the reader to \citet{PerezGonzalez24} and \citet{2025A&A...696A..57O} for details on the method, and for a discussion of improvements compared to ETC estimations. COSMOS-3D obtained data in the F1000W and F2100W filters, with exposure times of 927~s and 1848~s, respectively. The average $5\sigma$ detection limits for point-like sources measured in a circular aperture of radius $r=0.3, 0.7\arcsec$ for F1000W, F2100W are 24.8, 22.8~mag, respectively (calculated with the method described below). 

We measured photometry in the two MIRI bands assuming a point-like nature and applying aperture corrections based on the empirical PSFs provided by the \textit{JWST} Calibration Pipeline. We used several circular apertures in each band. We selected radii ranging from 0.2\arcsec\,  to 0.5\arcsec\, for F1000W and 0.4\arcsec\, to 1.0\arcsec\, for F2100W, the ranges of apertures sizes chosen to be able to detect faint sources and limited to maximum/minimum aperture losses of 50/25\% for point-like sources.  The background was measured in a $r=10\arcsec$ circular region around the source. For the background noise (used to get photometric uncertainties and the $5\sigma$ detection limits quoted above), following the method explained in \citet[][based on \citealt{2008ApJ...675..234P}]{2023ApJ...951L...1P},
we selected random non-contiguous pixels (separated by more than 3 pixels) to avoid the effects of noise correlation (i.e., an underestimation of the rms) introduced by the drizzling method when mosaicking the data.

For F1000W, we obtained consistent photometric measurements (within errors) with all apertures, the measurements ranging from $24.90\pm0.29$~mag for $r=0.2\arcsec$ to $25.63\pm0.59$~mag for $r=0.5\arcsec$. The final magnitude used in the rest of the paper, the one with maximum SNR obtained with $r=0.3\arcsec$ (aperture correction $0.60\pm0.03$~mag, the errors accounting for a 1~pixel centering error), is $25.02\pm0.26$~mag. 

For F2100W, we obtained negative fluxes or ${\rm S/N}<2$ measurements for all apertures, so we conclude a non-detection in this band, and  assume the $5\sigma$ upper limit quoted above of 22.8~mag (corresponding to an aperture of radius $r=0.9\arcsec$, where the encircled energy is $\sim70$\%).

\section{Data Analysis}\label{sec:analysis}
\subsection{Spectroscopic Redshift}\label{sec:redshift}

\begin{figure*}[ht]
\begin{flushright}
\includegraphics[width=0.94\textwidth]{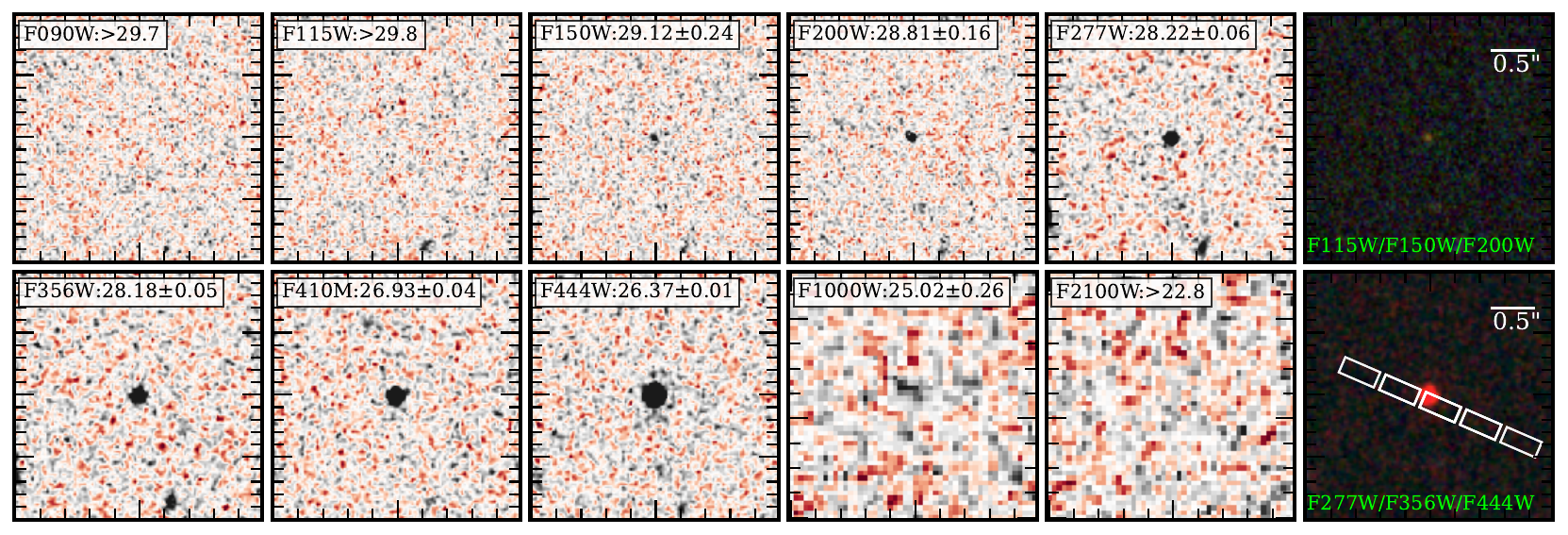}\\
\includegraphics[width=\textwidth]{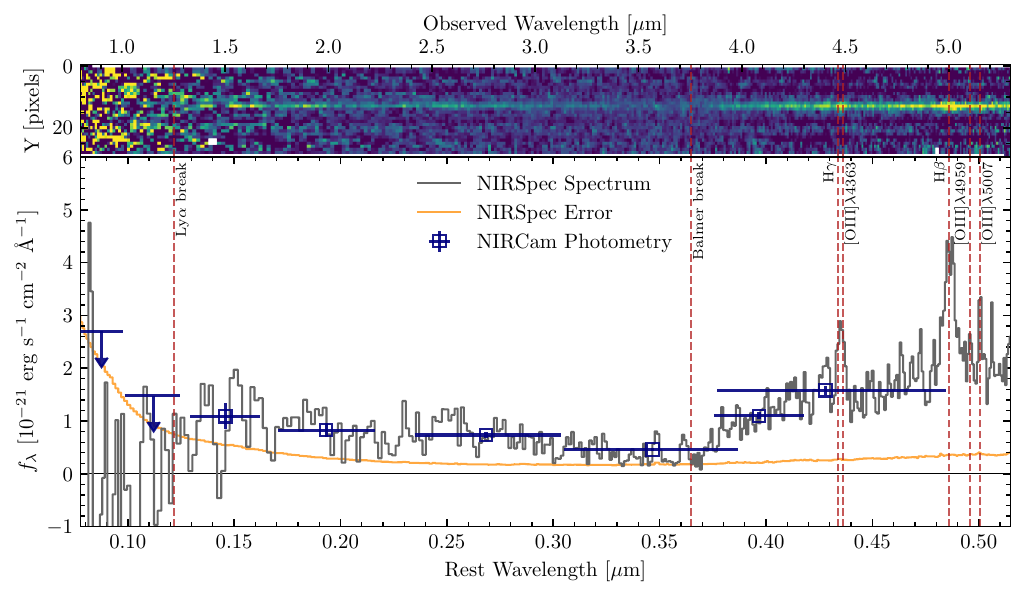}
\end{flushright}
\vspace{-12pt}
\caption{Top: $5'' \times 5''$ NIRCam and MIRI cutouts of \src{} in the PRIMER-COSMOS NIRCam and COSMOS-3D MIRI images. We also show RGB images generated for both the short wavelength (SW) and long wavelength (LW) NIRCam detectors. We overplot the NIRSpec MSA slitlet. Bottom: We plot the 2D and 1D photometrically calibrated spectra of \src{} in $f_{\lambda}$. We mark strongly detected emission features with red dashed lines and overplot the NIRCam photometry with blue squares. Despite being near the corner of the central shutter in the slitlet, \src{} exhibits a detectable \Lya break, a strong Balmer break, and clear rest-optical emission line detections--including broadened H$\beta$ and H$\gamma$, and narrow \oiii{}$\lambda\lambda 4959,5007$. The combination of narrow forbidden lines and a broad H$\beta$ clearly indicates that \src{} is a BLAGN at $ z=9.288$.}
\label{fig:fullspectrum}
\end{figure*}

The CAPERS team is using a combination of software tools, interactive inspection, and manual vetting to measure redshifts from the NIRSpec spectra; the full methodology will be detailed in a forthcoming publication. For \src{}, we derive the initial spectroscopic redshift using a modified version of \texttt{msaexp}\footnote{\url{https://github.com/VasilyKokorev/msaexp_OLF}} \citep{msaexp,kokorev24b}. Key modifications include the ability to vary the velocity width and to model individual lines with multiple components (e.g., narrow and broad).

We use \texttt{msaexp} to fit the continuum with cubic splines and emission lines with Gaussian profiles. For this step, we adopt \texttt{nsplines=20}, leaving the position, amplitude, and width of emission lines as free parameters. The full-width-half-maximum (FWHM) is allowed to vary between 150–800 km/s for narrow components and 800–5000 km/s for broad ones. Uncertainties are estimated via MCMC, using resampling of the covariance matrix. All fits are performed on spectra corrected to match the photometry. To account for the wavelength-dependent resolution of PRISM, the model grid is initially generated on an oversampled wavelength axis and then convolved with the instrument dispersion curve.

With this automated multi-line fitting approach we derive an initial $z_{\rm spec}=9.2880\pm0.065$ and a number of interesting features. Broad components in excess of $\sim3000$ km/s \textbf(FWHM) are required to adequately model both the H$\gamma$ and $H\beta$ emission, while the adjacent lines in the \oiii$\lambda\lambda4959,5007$ doublet remain narrow ($\sim 200$ km/s FWHM). The presence of broad permitted (e.g. Balmer series of hydrogen) and narrower (semi-) forbidden lines like \oiii, gives us a first hint that an accreting supermassive black hole is present in \src{}, typical of other spectroscopic investigations of LRDs with \textit{JWST} \citep[e.g.][]{matthee24,kocevski25,taylor24}. 

We note that this redshift-fitting routine is not always capable of accurately modeling all spectral features. While it performs well for most narrow and broad emission lines and is efficient for processing large spectral samples, its reliability decreases in the presence of complex line blends, uncommon line ratios, low signal-to-noise regions, or noisy continua. Accurate identification of key features, such as broad lines, is essential for robust classification of the object, especially when using the PRISM data. Therefore, in subsequent sections, we adopt a more refined fitting procedure for select regions of the spectrum.

\begin{figure*}[t]
\includegraphics[width=\textwidth]{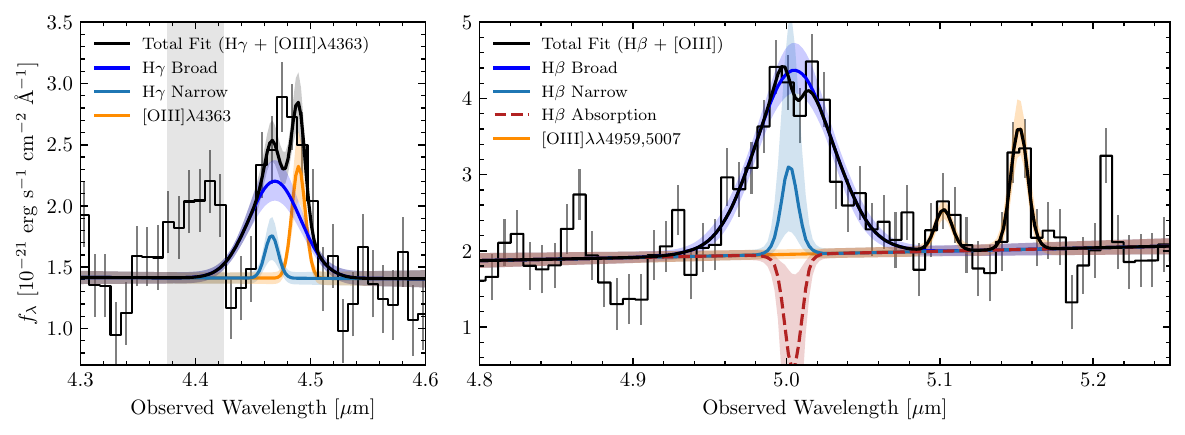}
\caption{The observed NIRSpec/PRISM spectrum of \src{} is shown as the fine black curve with uncertainties in gray.  Left: Line fits and 1$\sigma$ uncertainties on H$\gamma$ (blue curves and shaded regions), \oiii$\lambda4363$ (orange curve and shaded region) and the combined fit (thich black curve and shaded region). We de-blend these lines using the high confidence line widths and redshift solution measured from H$\beta$+\oiii$\lambda\lambda4959,5007$. We show the region of masked spurious pixels with gray shading. Right: Line fits to H$\beta$+\oiii$\lambda\lambda4959,5007$. A broad (FWHM$=3521\pm502$~km~s$^{-1}$, blue curve) component is clearly necessary to reproduce the H$\beta$. We also fit narrow emission (light blue) and absorption (dashed red) components to H$\beta$ to better model the double peak structure of the line, though their fluxes are degenerate at PRISM resolution. The \oiii$\lambda\lambda4959,5007$ doublet is also detected at ${\rm S/N}\sim4$. The combination of narrow \oiii$\lambda\lambda4959,5007$ and broad H$\beta$ are clear indicators of a broad-line AGN.}
\label{fig:linefits}
\end{figure*}

\subsection{Line Fitting}\label{sec:linefits}

The spectrum of our object presented in Figure \ref{fig:fullspectrum} coupled with our initial \texttt{msaexp} fit hint at the presence of broad component in both H$\gamma$ and H$\beta$ emission lines, in contrast to the narrower profiles of the adjacent \oiii$\lambda \lambda$4959,5007 \AA{} doublet. Further, the H$\beta$ line appears to be double peaked, which hints at an underlying absorption component. To evaluate the potential significance of these features we perform a more detailed line fitting to the H$\gamma$+\oiii$\lambda4363$ and H$\beta$+\oiii$\lambda\lambda4959,5007$ line complexes. The redder end of the PRISM where the H$\beta$+\oiii$\lambda\lambda4959,5007$ is located has a higher spectral resolution, and this line complex is not blended, therefore we fit that region first. We assume that the lines in the \oiii{} doublet are both narrow, while the H$\beta$ consists of a narrow, a broad and an absorption component. In our fitting, we assume that the velocity width of the narrow component for all three lines is the same, and we fix the ratio \oiii$\lambda5007$/\oiii$\lambda4959=2.98$ \citep{storey2000}.  Velocities of the narrow, broad and absorption components are allowed to vary between $50-800$, $1000-5000$ and $50-1000$ km s$^{-1}$, respectively. We fix the redshift (and thus the line centers of the narrow lines using the vacuum wavelengths from \citealt{vanhoof18}), however we allow for small ($\sim\pm500$ km/s) offsets in the absorption component and broad component of H$\beta$ (relative to the narrow \oiii) to allow for kinematic motion of gas. We model the local continuum with a first-order polynomial. 

We initialize the fit by first creating a set of models on the over-sampled wavelength grid. To mimic the variable resolution of the PRISM, we interpolate our model onto a variable step grid, while making sure that the total integrated flux is preserved. Early NIRSpec/MSA results \citep{degraaff23} have shown that the spectral resolution of a point-like source falling within a slitlet is higher, compared to a uniformly illuminated slit, sometimes up to a factor of two. We therefore conservatively increase the nominal spectral resolution by a factor of 1.7. To take into account the effects of the line spread function, we additionally convolve our model with Gaussians of variable resolution \citep{degraaff23,isobe23}. We then optimize this fit via MCMC with the \texttt{emcee} \citep{emcee} package for \texttt{python}. 

From our fit we securely (combined ${\rm S/N}\approx4.2$) confirm the presence of the \oiii$\lambda\lambda4959,5007$ doublet at a redshift $z=9.288\pm0.003$. We adopt this redshift for all subsequent calculations. We also detect a distinct broad component in H$\beta$ at ${\rm S/N}\approx10$, with an intrinsic ${\rm FWHM}=3521\pm502$~km~s$^{-1}$ and a tenuous slight redward velocity offset from the systemic redshift of $134\pm164$~km~s$^{-1}$. The width of the narrow lines is poorly constrained due to the low resolution of the PRISM, although we report a posterior value of ${\rm FWHM}=483\pm225$~km~s$^{-1}$. The narrow component of H$\beta$ does not appear to be statistically significant (${\rm S/N}\approx0.8$). Additionally, while including the H$\beta$ absorption component results in a visually better fit to the double peaked $H\beta$ line with a line width of ${\rm FWHM}=413\pm255$~km~s$^{-1}$, and slight redward offset of $84\pm126$~km~s$^{-1}$ this component is also statistically insignificant with (${\rm S/N}\approx1.1$) and highly degenerate with the narrow H$\beta$ component. These statistical non-detections may be expected at the resolution of the PRISM, but as they are physically motivated, we include them in the model regardless to better sample their covariance with the more significant model components. 

The situation with H$\gamma$+\oiii$\lambda4363$ is more complex, as these lines are blended at the PRISM resolution. We can however use the prior information obtained from the H$\beta$+\oiii$\lambda\lambda4959,5007$ complex and attempt to adequately separate the two lines. We fix the narrow and broad line widths, broad velocity offset, and redshift to the result we obtained from the H$\beta$+\oiii$\lambda\lambda4959,5007$ fit, and only fit the integrated fluxes of narrow+broad H$\gamma$, narrow \oiii$\lambda4363$, and a first-order polynomial continuum. Finally, the flux ratio of the narrow H$\gamma$ to narrow H$\beta$ is not allowed to exceed 0.47, as set by Case B recombination \citep{osterbrock89}. 
This fit is further complicated by six consecutive pixels at $4.375$--$4.425\mu$m that show flux above the level of the nearby continuum. While these pixels are detected at the $1$--$3\sigma$ level above the continuum, they correspond to no known emission line. Therefore, we conservatively mask out this spurious feature. We once again forward model the effects of the PRISM resolution and fit the H$\gamma$+\oiii$\lambda4363$ line complex with MCMC. Similarly to H$\beta$, we could not derive a significant flux for a narrow H$\gamma$ (${\rm S/N}\approx2$), however, we detect and partially deblend a broad H$\gamma$ component at ${\rm S/N}\approx4.8$ and \oiii$\lambda4363$ at ${\rm S/N}\approx3.1$. We note that the blue wing of the broad H$\gamma$ line appears higher than the data.  This may indicate that the masked spurious feature is affecting these pixels, or that the H$\gamma$ broad line width is narrower than that of H$\beta$, as has been seen in low-redshift quasars \citep[e.g.][]{bentz23}. As our analyses are not dependent on the H$\gamma$ width, this uncertainty does not affect our interpretation of this object.  We present various measured and derived properties of \src{} in Table~\ref{tab:linefits} and show our best-fit models to both line complexes in Figure~\ref{fig:linefits}.

\begin{deluxetable}{ll}
\label{tab:linefits}
\caption{Properties of \src{}}
\tablehead{Property & Value}
\startdata
R.A. & 150.1362532 deg\cr
Decl. & 2.3080298 deg\cr
Redshift ($z$) & $9.288\pm0.003$ \cr \hline
H$\gamma$ flux (narrow) & $<7.1$ \cr 
H$\gamma$ flux (broad)  & $45.9\pm9.5$ \cr
\oiii{}$\lambda$4363 flux  & $13.0\pm4.2$ \cr 
H$\beta$ flux (narrow) & $<62.7$\cr 
H$\beta$ flux (broad) & $140\pm14.6$ \cr
H$\beta$ flux (absorption)  & $>-69.1$~\cr 
H$\beta_{\rm broad}$ FWHM & $3521\pm502$~km~s$^{-1}$\cr
H$\beta_{\rm abs}$ FWHM & $282\pm159$~km~s$^{-1}$\cr
H$\beta_{\rm broad}$ offset & $134\pm164$~km~s$^{-1}$\cr
H$\beta_{\rm abs}$ offset & $84\pm126$~km~s$^{-1}$\cr
\oiii{}$\lambda$4959 flux  & $7.2\pm3.5$ \cr
\oiii{}$\lambda$5007 flux  & $22.2\pm5.3$ \cr 
\oiii{}$\lambda$5007 FWHM & $483\pm225$~km~s$^{-1}$\cr \hline
$M_{1500\textrm{\AA{}}}$ & $-18.2\pm0.2$~mag \cr 
$\beta_{\rm UV}$ & $-0.99^{+0.14}_{-0.13}$ \cr \hline
$\log\left(M_{\textrm{BH}}/M_{\odot}\right)$ (canonical) & $7.58\pm0.15$ \cr
$\log\left(M_{\textrm{BH}}/M_{\odot}\right)$ (systematic bounds) & $6.65$--$8.50$ \cr
$A_{\rm V,BLR}$ & $1.9^{+1.3}_{-1.2}$ \cr
$\log\left(M_{*}/M_{\odot}\right)$ & $<9.0$\cr
$M_{\textrm{BH}}/M_*$ (canonical) & $>4.5\%$ \cr
$M_{\textrm{BH}}/M_*$ (systematic bounds) & $>46\%,>0.5\%$ \cr
Balmer Break: $\left(f_{\nu,4050\textrm{\AA{}}}/f_{\nu,3670\textrm{\AA{}}}\right)$ & $4.35^{+0.93}_{-0.67}$ \cr
\enddata
\tablecomments{All line fluxes are given in units of $10^{-20}$~erg~s$^{-1}$~cm$^{-2}$.  Lines fit at S/N$\leq$2  are listed as 2$\sigma$ (upper/lower) limits on their fluxes.  We define a ``canonical'' $\log\left(M_{BH}/M_{\odot}\right)$ and uncertainties as only those errors propagating from the measured broad-line properties and the empirical uncertainties given in \citep{greene05} without correcting for dust. We further provide upper and lower bounds on $\log\left(M_{\textrm{BH}}/M_{\odot}\right)$ to account for systematic uncertainties in the measurement methods (see \S\ref{sec:bhmass} for details).}
\end{deluxetable}

\subsection{Morphology and Size Measurements}\label{sec:morphology}

\begin{figure}
\includegraphics[width=\columnwidth]{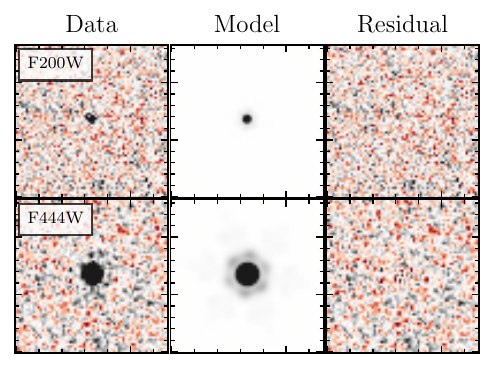}
\caption{Results of two-dimensional surface brightness profile fitting. We show $2^{\prime\prime}\times2^{\prime\prime}$ cutouts in F200W and F444W around \src{} in the left column. Our best-fit, point-source models in each band are shown in the middle column, while the residuals (data-model) are shown in the rightmost column. \src{} is unresolved in all bands, with $r_{h} < 0\farcs04$ and $0\farcs08$ in F200W and F444W, respectively, corresponding to respective physical sizes of $\lesssim 175$~pc and $\lesssim 350$~pc.}
\label{fig:galfit}
\end{figure}

To determine if any of the emission from \src{} might originate from an extended stellar population, we use the \texttt{GALFIT} software \citep{peng02} to model the galaxy as a point source in several NIRCam bands, to search for signs of an underlying galaxy in the residual image. For this modeling, we provide \texttt{GALFIT} with empirical PSFs constructed from the PRIMER-COSMOS mosaic and noise images that account for both the intrinsic image noise (e.g., background and readout noise) and added Poisson noise due to the objects themselves. We show the results of this morphological modeling in Figure~\ref{fig:galfit} for F200W and F444W.  We see no signs of any extended structure, indicating that the source is unresolved or the surface brightness of the resolved component is too low for detection with the sensitivity of our image.  
We estimate an upper limit on the size of the host to be $r_{h} < 0\farcs04$ and $0\farcs08$ in F200W and F444W, respectively, which correspond to physical sizes of $\lesssim 175$ pc and $\lesssim 350$ pc.

\subsection{Spectrophotometric Fitting}\label{sec:templates}

\begin{deluxetable}{@{\extracolsep{10pt}}l@{}c@{}c@{}}
\caption{\texttt{Cloudy} parameter grid \& fitted values}\label{tab:models}
\tablehead{Parameter & Grid Values & Fitted Value}
\startdata
$\log T_{\rm BB}$/K & $4$, $4.7$, $5$, $5.7$ & $5.0$ \cr
$\alpha_{\rm OX}$ & $-2.5$, $-2.0$, $-1.5$ & $-1.5$ \cr
$\alpha_{\rm UV}$ & $-0.1$ & \dots \cr
$\alpha_{\rm X}$ & $-0.5$ & \dots \cr
$\log n_H/{\rm cm}^{-3}$ & $9, 9.5, 10, 10.5, 11, 11.5, 12$ & $9.9^{+0.2}_{-0.2}$ \cr 
$v_{\rm turb}/{\rm km}\,{\rm s}^{-1}$ & $100, 200, 300, 400, 500$ & $320^{+80}_{-60}$ \cr
[Fe/H] & $-2$ & \dots \cr
$\log U$ & $-3.5, -3, -2.5, -1.5, -0.5$ & $-1.5$
\cr
$\log N_H/{\rm cm}^{-2}$ & $21, 22, 23, 24, 25, 26$ & $>25.9^\dagger$ \cr
\hline 
$C_f$ & 0--1 & $0.12^{+0.01}_{-0.01}$ \cr 
$A_V$ & 0--3 (SMC law) & $0.53^{+0.09}_{-0.08}$
\enddata
\tablecomments{The first four parameters describe the incident continuum spectrum where $T_{\rm BB}$ is the `Big Bump' temperature, $\alpha_{\rm OX}$ is the optical to X-ray index, $\alpha_{\rm UV}$ is the power-law UV slope and $\alpha_{\rm X}$ is the power-law X-ray slope. The incident continuum is passed through dense gas defined by the gas density ($n_H$), metallicity ([Fe/H]) and turbulent velocity ($v_{\rm turb}$). The level of irradiation at the face of the cloud is set by the ionization parameter ($\log U$) and the \texttt{Cloudy} calculations are stopped at a range of line-of-sight column densities ($N_H$). \\
$^\dagger$\,We provide the 1$\sigma$ lower limit for the column density as our posterior is limited by the edge of the \texttt{Cloudy} grid.}
\end{deluxetable}

We model the SED of \src{} as a composite galaxy+AGN, similar to previous work on LRDs with Balmer breaks \citep[e.g.][]{wang24,labbe24}. However, here we model the Balmer break as part of the AGN continuum (rather than old stars), driven by absorption from a shell of dense gas around the accretion disk/broad line region \citep[BLR; e.g.][]{inayoshi25, ji25, naidu25}. 
We explored modeling the spectrum using only a stellar component, with flexible non-parametric star-formation history (SFH) driving the strong Balmer break with an old stellar population ($>300$ Myr). 
Comparing the fluxes on either side of the break (in 100\,\AA-wide windows centered at 3670 and 4050\,\AA), our best-fit stellar-only model has a break strength of $\sim 2.8$.
This is insufficient to match the strength of the break in \src{} (which we measure as $f_{\nu,4050\textrm{\AA{}}}/f_{\nu,3670\textrm{\AA{}}}=4.35^{+0.93}_{-0.67}$), similar to other recently discovered LRDs with extreme Balmer breaks (RUBIES-BLAGN-1; $2.73^{+0.20}_{-0.17}$; \citealt{wang24a}, Abell2744-QSO1; $3.33^{+0.15}_{-0.15}$; \citealt{weibel25,ma25}, MoM-BH*-1; $7.7^{+2.3}_{-1.4}$; \citealt{naidu25}, RUBIES-UDS-154183; $6.9^{+2.8}_{-1.5}$; \citealt{degraaff25}).
The stellar-only model is overall a poor fit to the data, with a Bayesian Information Criterion (BIC) difference from our fiducial model (described below) of $\Delta{\rm BIC} > 80$. 
This, combined with the significant detection of broad H$\beta$, drives our assumption that the rest-frame optical is dominated by the AGN, enshrouded in dense gas.

We therefore pursue a more complex modeling, first using the \texttt{Cloudy} photoionization code (version 23.01; \citealp{cloudy23}) to generate a large grid of AGN model SEDs, varying both the intrinsic AGN accretion disk SED and the gas conditions around the source. We follow the framework described in \cite{inayoshi25} and \cite{naidu25} and model the intrinsic AGN accretion disk SED as a series of power laws combined with a `Big Bump' temperature.\footnote{This AGN continuum model is described in Section 6.2 of the HAZY1 \texttt{Cloudy} documentation (version 23.01).} This continuum is passed through a shell of gas surrounding the central source which is defined in terms of its density, metallicity, and turbulent velocity. The turbulence is required to reproduce the break shape; without it, the break is very sharp at $\lambda = 3650$\,\AA. We vary the ionization parameter which sets the ratio of H-ionizing photons to gas densities at the illuminated face of the gas cloud. When all other parameters are fixed, varying the ionization parameter is equivalent to varying the distance between the central source and the gas shell. We stop the calculation at a set of fixed line of sight column densities through the gas cloud. The full grid of \texttt{Cloudy} model parameters is provided in Table~\ref{tab:models}. 

In order to conduct joint inference on the AGN and galaxy parameters, we assemble and fit the full SED model using a modified version of \texttt{bagpipes} \citep{carnall18}, a Bayesian SED modeling code. 
\texttt{bagpipes} requires models with a continuous parameter space, rather than discrete points in a grid; we include a custom module to interpolate over the \texttt{Cloudy} grid (consisting of both the continuum and emission  lines of the AGN processed through the dense gas shell). 

However, in an effort to reduce the dimensionality of the problem, prior to fitting with \texttt{bagpipes} we first perform a simple $\chi^2$ minimization routine over the \texttt{Cloudy} grid to identify and fix best-fit values of several grid parameters. 
Here, we fit only to the PRISM spectrum redward of the Balmer break, and include several post-processing steps, including broadening of the AGN emission lines, a variable covering fraction of the BLR clouds, and dust attenuation following an SMC law \citep{gordon03}.
Fitting to our grid, we find the best-fit model has $\log T_{\rm BB}/K = 5$, $\alpha_{\rm OX} = -1.5$, $\log n_H/{\rm cm}^{-3} = 9.5$, $v_{\rm turb} = 100$ km s$^{-1}$, $\log U = -1.5$, and $\log N_H/{\rm cm}^{-2} = 26$. We note that these parameters describe a Compton-thick environment which \texttt{Cloudy} is not designed to simulate\footnote{See Section 3.4 of the HAZY2 \texttt{Cloudy} documentation}. Therefore, while we caution that our \texttt{Cloudy} models may not capture all physical processes in this regime, \texttt{Cloudy} remains the best accessible tool for this modeling.

In the subsequent modified \texttt{bagpipes} fit, we then fix $T_{\rm BB}$, $\alpha_{\rm OX}$, and $\log U$ at their best-fit values and interpolate over the density $\log n_H$, column density $\log N_H$, and turbulent velocity $v_{\rm turb}$. 
These parameters predominantly impact the strength and shape of the Balmer break \citep[see][]{ji25}, and are therefore important to model flexibly to optimize the fit. 
We adopt uniform priors for each parameter, and again include a broadening of the AGN emission lines, a variable covering fraction of the BLR clouds ($C_f$), and dust attenuation ($A_V$; SMC law); we note that we explored allowing the attenuation curve slope to vary, but find no need for an extremely steep dust law \citep[in contrast to][]{degraaff25}. 

For the galaxy model, we use the BPASS v2.2.1 stellar templates \citep{bpass221} and a flexible star-formation history via the ``bursty continuity'' parametrization described in \citet{tacchella22}. We use five fixed age bins with edges at 0, 10, 30, 100, 200, and 300 Myr and a log-uniform prior on the stellar metallicity from 0.1\% to 100\% solar. We include nebular emission, varying $\log U$ from $-4$ to $-1$, and dust attenuation (SMC law). Note that the dust attenuation is fit separately for the stars and AGN. 

\begin{figure*}[ht]
\includegraphics[width=\textwidth]{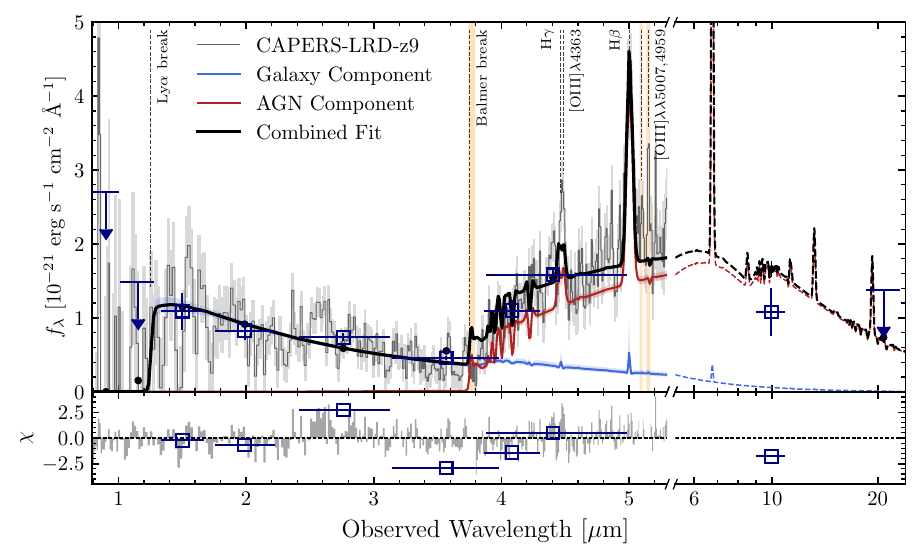}
\caption{Spectrum and 1$\sigma$ errors of \src{} (gray curve, light gray shading), the best fit host galaxy component (blue and dashed blue curves), the best fit dense-gas enshrouded AGN component (red and dashed red curves), combined host+AGN fit (black and dashed black curves), photometry data (dark blue squares and upper limits), best-fit model photometry (black points) in the upper panel, and masked regions (gold shading). The lower panel shows the $\chi$ residuals of the fit for the spectrum (gray curve) and photometry (dark blue squares). Note that neither the \texttt{BAGPIPES} stellar model nor the \texttt{Cloudy} AGN model can match the strong \oiii{}$\lambda$4363 emission.}
\label{fig:combinedfit}
\end{figure*}

The final model is the combination of the stellar and AGN components, with the normalization of the latter fit as a free parameter. 
The model spectrum is convolved with the PRISM resolution curve and the joint model is fit to both the NIRSpec spectrum and the NIRCam+MIRI photometry. 
We note that we mask the \oiii{}$\lambda\lambda$4959,5007 doublet when fitting the spectrum, as the narrow-line region (NLR, see \S\ref{sec:temden})---which is not considered in our \texttt{Cloudy} modeling of the BLR and nearby gas or our stellar model---may significantly contribute to these lines. 
We additionally mask the region around H$_\infty$ (3630--3690\,\AA), where the \texttt{Cloudy} modeling yields artifacts due to the finite number of resolved energy levels \citep{inayoshi25,ji25}. 
The resulting best-fit model is shown in Figure~\ref{fig:combinedfit}, and the results from the spectrophotometric modeling are discussed in Sections~\ref{sec:stellarmass} and \ref{sec:temden}.

\section{Results}\label{sec:results}

\subsection{BLR Dust Attenuation}\label{sec:dust} 

The ratio of observed fluxes from the Balmer series lines provides a measure of dust attenuation. Here we use the observed H$\beta$/H$\gamma$ ratio. 
As illustrated in Figure \ref{fig:linefits}, the H$\gamma$ line is blended with \oiii$\lambda4363$. While we use priors based on the H$\beta$+\oiii$\lambda\lambda4959,5007$ line widths to de-blend these lines, the per-pixel S/N for H$\gamma$ is not sufficient for a double-component fit. Since we cannot isolate the narrow component of H$\gamma$ or H$\beta$, we compute a Balmer decrement for the BLR solely based on the ratio of H$\beta$ and H$\gamma$ broad components' fluxes, ensuring flux integration across comparable velocity ranges. This yields H$\beta$/H$\gamma=3.5^{+1.7}_{-0.9}$.

To calculate $A_{\rm V}$, we apply the Small Magellanic Cloud (SMC) reddening law \citep{gordon03}, which has been found to match well the dust attenuation in high-$z$ galaxies \citep{capak15,reddy15,reddy18} and reddened quasars \citep[e.g. see][]{hopkins04}. 
Assuming Case B recombination, the intrinsic line ratio would be (H$\beta$/H$\gamma$)$_{\rm int}$=2.14 \citep{osterbrock89}. Our observed ratio suggests significant attenuation; we estimate $A_{\rm V,BLR}=1.9^{+1.3}_{-1.2}$ (where our precision is limited primarily by the ${\rm S/N}\approx4$ detection of broad H$\gamma$). We note that while Case B recombination may not be a sound assumption for collisionally excited gas, this nonetheless provides an estimate of the reddening of the BLR.

\subsection{Black Hole Mass}\label{sec:bhmass}

Black hole mass estimation from broad emission lines relies on two fundamental assumptions: the broad-line emitting gas is virialized and dominated by the gravitational potential of the central black hole, and a ``radius--luminosity'' relationship exists between the broad-line orbital radius and the AGN luminosity (in Equation 1, the H$\beta$ broad-line luminosity). These assumptions have generally been validated for low-redshift AGN \citep[e.g.][]{Bentz2006, Cackett2021} but their applicability to the broader AGN population is less certain. 

In the specific case of \src{}, the broad and symmetric H$\beta$ line profile is consistent with the basic virial assumption for kinematics dominated by gas orbits around a massive black hole, in contrast to the asymmetric broad-line profiles commonly observed in quasars with significant non-virial kinematics (e.g. \citealt{U2022, Fries2024}). The reliability of the radius--luminosity assumption is more difficult to assess. Compared to the radius--luminosity relation measured for Seyfert~1 AGN that was used to calibrate single-epoch masses, recent studies have shown that luminous and rapidly accreting quasars have smaller BLR sizes \citep{Du2016, FonsecaAlvarez2020}, which leads to smaller black hole masses. The empirical radius--luminosity relation is generally assumed to be caused by a photoionization-bounded broad-line region \citep[e.g.][]{Korista2004}, and the unusual SED shape of LRDs like \src{} could result in a significantly different BLR structure compared to other quasars. Elevated Eddington ratios also affect the radius--luminosity relation and may also result in the overestimation of black hole masses by as much as an order of magnitude \citep[e.g..][]{Du2016,lupi24}. We ultimately use the canonical relationship for low-redshift AGN of Equation~\ref{eq:bhmass} to estimate a mass for \src{}, while acknowledging that significant work must be done to better understand the applicability of mass estimates for high-redshift LRDs.

To compute the canonical black hole mass, we use Equation~10 from \cite{greene05}, where $L_{{\rm H}\beta}$ and $\textrm{FWHM}_{{\rm H}\beta}$ are the luminosity and FWHM of the broad component of H$\beta$, respectively. 
\begin{equation} \label{eq:bhmass}
\begin{aligned} 
M_{\textrm{BH}}&=(2.4\pm0.3) \\ & \times 10^6\left(\frac{L_{{\rm H}\beta}}{10^{42}\textrm{ erg s}^{-1}}\right)^{0.59\pm0.06} \\ & \times \left(\frac{\textrm{FWHM}_{{\rm H}\beta}}{10^3\textrm{ km s}^{-1}}\right)^{2}M_{\odot} 
\end{aligned}
\end{equation}
Propagating the uncertainties on the calibration coefficients given in the empirical relation and the uncertainties on our line flux and FWHM, we derive a black hole mass for \src{} of $\log(M_{\textrm{BH}}/M_{\odot})=7.58\pm0.15$. We note that additional systematic uncertainties from applying the \cite{greene05} relations at high redshift may be as high as 0.5-0.7~dex \citep{FonsecaAlvarez2020,abuter24}.

To best capture these true systematic uncertainties and account for the limitations of our PRISM data, we next compute strong upper and lower limits on $\log(M_{\textrm{BH}}/M_{\odot})$.
First, we derive an upper bound by correcting our canonical measurement for dust attenuation. By adopting the estimated reddening of $A_{\rm V,BLR}=1.9^{+1.3}_{-1.2}$ (computed above), we derive a higher black hole mass of $\log(M_{\textrm{BH}}/M_{\odot})=8.10\pm0.40$. Using our $A_V\approx0.53$ from our spectrophotometric fitting, we alternatively compute $\log(M_{\textrm{BH}}/M_{\odot})=7.70\pm0.14$. We note that while our Balmer decrement based- and spectrophotometric fitting based-$A_V$ values are discrepant at the 1.3$\sigma$ level, we attribute this difference to the uncertain assumption of Case B recombination in the BLR, uncertainty on the H$\gamma$ line flux due to the limitations of the PRISM resolution, and possible collisional de-excitation of neutral in the dense-gas region of the LRD. Nonetheless, we adopt the 1$\sigma$ upper limit on the Balmer decrement derived value ($\log(M_{\textrm{BH}}/M_{\odot})<8.5$) as a strong upper limit on the black hole mass.

We also compute a lower limit on the black hole mass using the H$\beta$ line luminosity and assuming that the AGN is accreting at the Eddington rate. For the purposes of computing a lower limit, we conservatively convert the H$\beta$ luminosity to H$\alpha$ luminosity assuming an unattenuated Case B ratio (H$\alpha$/H$\beta$=2.86; \citealt{osterbrock89}), and use the bolometric correction from \cite{stern12} to derive $\log(M_{\textrm{BH}}/M_{\odot})>6.65$. This method has the benefit of being insensitive to the measured H$\beta$ line-width. As such, it remains robust even if the H$\beta$ line-width is significantly overestimated due to uncertainties in the effects of the NIRSpec/PRISM spectral resolution, or more exotic explanations for line broadening such as electron scattering \citep[e.g.,][]{rusakov25}. 

When considering these upper and lower limits together, we compute systematic bounds on $\log(M_\textrm{BH}/M_\odot)$ of $6.65$--$8.50$. However, for ease of comparisons to other reported LRD black hole masses in the literature, we adopt our canonical measurement and its associated non-systemic error bounds for the remainder of this work (unless otherwise stated).  

\subsection{Stellar Mass}\label{sec:stellarmass}

Properly estimating the stellar masses of LRDs has proven enormously difficult given the unclear origin of the ``v-shaped'' spectrum and degeneracies in decomposing the galaxy/AGN contribution \citep[e.g.][]{kokorev23, Barro23, furtak24, wang24a, wang24, akins24, labbe24}. 
Here, we have assumed that the rest-UV is dominated by stars (with the rest-optical dominated by the AGN); however, we do not know this {\it a priori}. 
The stellar mass we derive is likely therefore an upper limit, as it is entirely possible that the UV is also AGN dominated (a possibility we discuss further in \S\ref{sec:nonstellar}). 
The joint \texttt{bagpipes}+\texttt{Cloudy} fitting described in \S\ref{sec:templates} yields a host galaxy stellar mass of $\log(M_*/M_{\odot})=8.9^{+0.1}_{-0.1}$.

We note that stellar masses can be underestimated when derived from the UV alone, due to ``outshining'' from young O/B stars \cite[see e.g.][]{narayanan24}.
However, as there are no prominent emission features in the rest-UV of \src{}, and the rest-optical emission lines are accounted for via the AGN, the SFH model is not biased towards a young stellar population. 
In fact, the age-dust-metallicity degeneracy is accounted for in our posterior estimate. 
We explored including an additional burst of SF at $z=20$, and found a consistent stellar mass with no meaningful improvement to the quality of the fit, showing that the observations do not allow for a significant mass in older stars to be present in this object. 

\begin{figure}[t]
\includegraphics[width=\columnwidth]{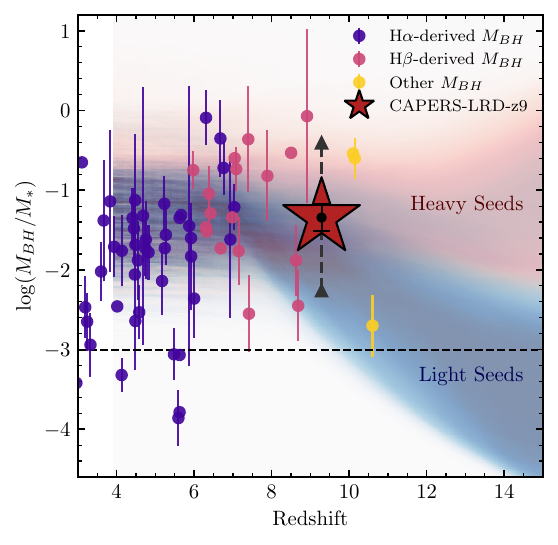}
\caption{Black hole mass to stellar mass ratio $\log\left(M_{\textrm{BH}}/M_*\right)$ as a function of redshift for \src{} (red star) and a collection of literature \textit{JWST}-detected AGN \citep{harikane23,kocevski23,kokorev23,larson23,bogdan24,furtak24,kokorev24,labbe24,maiolino24,napolitano24ghz9,tripodi24,ubler24,wang24a,yue24eiger,akins25,kocevski25,juodzbalis25}, showing the published errors, typically exclusive of systematic uncertainties. We divide the literature selected samples into those selected based on a broad H$\alpha$ line (purple points), a broad H$\beta$ line (pink points), and based on UV line or X-ray detections (yellow points). We plot the statistical errors on \src{} as the black error bars, and we include a larger systematic range on $\log\left(M_{\textrm{BH}}/M_*\right)$ based on the systematic limits on the black hole mass of \src{} as a gray dashed line and upward pointing arrows. We stress that as our $M_*$ measurement is an upper limit, all of our measurements of $M_{\textrm{BH}}/M_*$ are lower limits. We plot the $M_{\textrm{BH}}/M_*=0.1\%$ value for local galaxies as a black dashed line. For comparison, we overlay the predicted evolution of $M_{\textrm{BH}}/M_*$ ratios from early BH populations formed via two distinct seeding channels: heavy seeds (red shaded region) and light seeds (blue shaded region), based on the semi-analytical model presented in \citet{Hu_BHMF2025}. The population of \textit{JWST}-detected AGN exhibits an apparent over-massiveness in this ratio compared to the local relation.}
\label{fig:mbhmstar}
\end{figure}

From our computed stellar mass upper limit and our canonical measurement of $M_{\textrm{BH}}$, we compute $M_{\textrm{BH}}/M_*>4.5\%$. Invoking our systematic lower and upper bounds for $M_{\textrm{BH}}$, we alternatively compute $M_{\textrm{BH}}/M_*>0.5\%$ and $M_{\textrm{BH}}/M_*>46\%$, respectively. 
We plot this black hole to stellar mass ratio in Figure~\ref{fig:mbhmstar}, along with a large body of literature samples \citep{akins25,furtak24,harikane23,juodzbalis25,kocevski23, kocevski25,kokorev23,kokorev24,maiolino24,labbe24,larson23,tripodi24,ubler24,wang24a,yue24eiger}. 
Figure~\ref{fig:mbhmstar} demonstrates the ``over-massiveness'' of \src{} in the context of both literature samples of \textit{JWST}-detected AGN (colored points) as well as the established mass ratio for local quasars of $0.1\%$ (dashed black line). 
For comparison, we overlay the predicted evolution of $M_{\textrm{BH}}/M_*$ ratios from $\sim 10^3$ early BH populations formed via two distinct seeding channels: heavy seeds (red contours) and light seeds (blue contours), based on the SP1 model presented in \citet{Hu_BHMF2025}.

The observed value of for \src{} is broadly consistent with both scenarios, but the lower limit nature of this value may hint toward the heavey seed regime.
We discuss possible explanations for this effect and its possible interplay with black hole seeding and growth in \S\ref{sec:bhgrowth}.

\subsection{Gas density \& temperature}\label{sec:temden}

Our best-fit AGN model has $\log n_H/{\rm cm}^{-3} \sim {9.9}^{+0.2}_{-0.2}$, and $\log N_H/{\rm cm}^{-2} \gtrsim 26$.\footnote{Note that these results describe a Compton-thick regime that \texttt{Cloudy} is not explicitly designed to model (see \S\ref{sec:templates}).}
These are similar to the best-fitting parameters for MoM-BH*-1 and RUBIES-UDS-154183 \citep{naidu25, degraaff25}, and suggest very extreme gas conditions in the BLR. 
To examine whether these conditions extend to the NLR, we look to the forbidden oxygen lines. 
We measure an extremely high ratio of \oiii{}$\lambda4363$/\oiii{}$\lambda5007$ (RO3) $= 0.58\pm0.23$ in \src. 
Though \oiii{}$\lambda$4363 is blended with H$\gamma$ at the PRISM resolution, our MCMC line fitting (described in \S\ref{sec:linefits}) yields a significant \oiii{} detection (SNR $\sim 3$).

As the \oiii{}$\lambda$4363 and $\lambda$5007 lines arise from different upper energy levels, RO3 is commonly used as a temperature diagnostic. 
High RO3 traces high temperature regions of the ISM, and is found to be increasingly common in $z\gtrsim 8$ galaxies \citep{brinchmann23,katz23b,cullen25}. 
RO3 has also been used as an AGN diagnostic, particularly at high-redshift, where the canonical BPT diagram breaks down due to lower metallicity \citep{backhaus23, mazzolari24}. 
Here, the \oiii{}4363 line likely traces the NLR, or ISM gas heated by the AGN in addition to star-formation. 
 
The ratio we observe for \src{} ($\mathrm{RO3} \approx 0.58\pm0.23$) is particularly extreme, even compared to other $z>7$ BLAGN \citep{kokorev23,tripodi24,ubler24}.
Such a high ratio is only possible at densities approaching (or exceeding) the critical density of \oiii{}$\lambda$5007 ($n_e \sim 7\times 10^5\,{\rm cm}^{-3}$) where collisional de-excitation becomes important.
The critical density for \oiii{}$\lambda$4363 is significantly higher ($\sim 10^8\,{\rm cm}^{-3}$). 
Using \texttt{pyneb} \citep{luridiana15}, we find that RO3 we measure corresponds to densities $\gtrsim 10^6$ cm$^{-3}$, even at very high temperatures $T_e\sim 10^5$ K. 
At lower temperatures ($T_e \sim 10^4$ K), the implied density is even higher: $n_e\gtrsim 10^8$ cm$^{-3}$. 
This suggests that the NLR itself could be quite dense, corroborating the extreme densities inferred for the BLR. 

\section{Discussion}\label{sec:discussion}

\subsection{Implications for Black Hole Seeding and Growth}\label{sec:bhgrowth}

\begin{figure*}[ht]
\includegraphics[width=\textwidth]{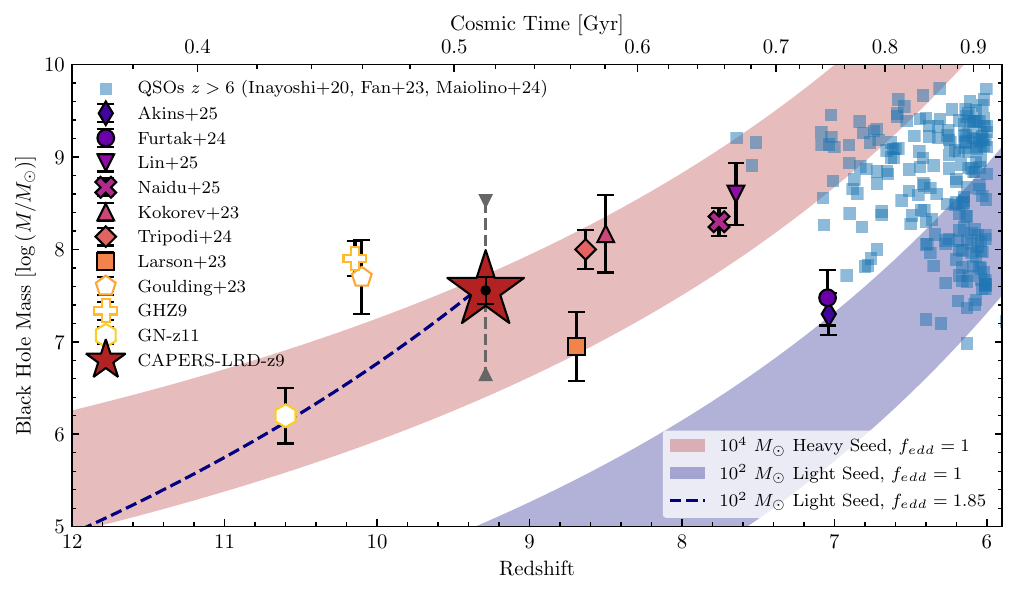}
\caption{Redshifts and black hole masses of \src{} (red star; canonical errors are shown in black, systematic upper and lower limits are shown in gray), populations of $z\gtrsim6$ quasars (small blue points, \citealt{Inayoshi2020,fan23,maiolino24}), notable massive $z>6$ spectroscopically confirmed BLAGN, (solid symbols, \citealt{akins25,furtak24,lin25,naidu25,kokorev23,tripodi24,larson23}), and the highest redshift AGN detected through X-ray emission and high-ionization UV emission lines, respectively (unfilled symbols, \citealt{goulding23,bunker23,kovacs24,napolitano24ghz9}). We also show the growth of $10^2 M_{\odot}$ (blue shading) and $10^4 M_{\odot}$ (red shading) black hole seeds growing at the Eddington limit. 
We also show the growth track of a $10^2 M_{\odot}$ stellar remnant formed at $z=30$ that starts accreting at $1.85\times$ the Eddington limit 100~Myrs after formation (dark blue dashed curve). \src{}'s black hole is too massive to be the result of an Eddington-limited stellar seed, thus a stellar remnant light seed undergoing periods of super-Eddington accretion or a heavy seed are necessary to produce \src{}'s black hole by $z=9.288$.}
\label{fig:bh_growth}
\end{figure*}

We plot our canonical black hole mass for \src{} in Figure~\ref{fig:bh_growth}, along with other notable $z>6$ AGN from the literature. We also plot simple models of Eddington-limited black hole growth (with 10\% radiative efficiency) for both stellar (light) seeds ($\sim10^2 M_{\odot}$) and more massive heavy seeds ($\gtrsim10^4 M_{\odot}$). The red shaded region corresponds to heavy seeds with formation redshifts of 25--15 that begin accreting at the Eddington limit immediately after formation. For the stellar remnant light seeds, we plot formation redshifts of 30-15 and require an additional 100~Myr to pass after the formation redshift before accretion begins due to progenitor gas heating by the pre-collapsed stars \citep[e.g.,][]{Johnson2007}. It is clear that \src{}'s SMBH is too massive to form from a simple stellar remnant that grows at the Eddington rate, similar to what has been inferred for other recently discovered high-$z$ AGN \citep[e.g.,][]{larson23}. However, a heavy seed growing at the Eddington rate can easily reproduce the observed BH mass, making it a plausible formation pathway. This scenario naturally establishes the over-massive $M_\textrm{BH}/M_*$ configurations present at high-$z$, as shown in Figure~\ref{fig:mbhmstar} (see also the red shading representing the heavy-seed model prediction).
Furthermore, the emerging SMBH demographics suggest that the BH-to-stellar mass ratio may bifurcate as we approach the highest redshifts, into a high- and low-$M_{\rm BH}/M_{\ast}$ branch. Such a bifurcation has been predicted for a hybrid seeding scenario, where the two branches are linked to heavy and light seeds, respectively (also see Figure~3 in \citealt{Jeon_BHMF2025}).

A key uncertainty in all direct-collapse BH (DCBH) models is the choice of formation criteria, such as the required level of soft-UV (LW) background flux, or the critical metallicity above which gas cooling would become too efficient, resulting in cloud fragmentation which suppresses DCBH formation \citep[e.g.][]{Bromm_Loeb_2003,Begelman2006,omukai08,Wise_DCBH2019}. The value of these parameters is reflected in the resulting number densities of DCBH seeds and their descendants \citep[e.g.,][]{Trinca_CAT2022,Hu_BHMF2025,Jeon_BHMF2025}. In turn, any firm limits on the DCBH abundance would significantly constrain the formation scenarios, as could be accomplished by linking the DCBH seeding pathway to a subset of the observed LRD population. 

Some studies have suggested that the conditions required to form DCBH seeds are too rare for this channel to account for the majority of the SMBHs being detected by JWST \citep[e.g.,][]{Bhowmick_SMBH2024}, whereas others argue for less restrictive conditions \citep[e.g.,][]{Chon_DCBH2025}. However, observations suggest that ultradense star clusters may be common or even ubiquitous in high redshift galaxies \citep{Adamo2024,Mowla2024,Fujimotograpes2024}, providing a promising additional heavy seeding mechanism.

Furthermore, the heavy seed explanation is not the only way to produce \src{}'s black hole mass by $z=9.288$. Allowing (even mild) super-Eddington accretion with a stellar remnant seed can also reproduce the black hole mass of \src{}. For example, in Figure~\ref{fig:bh_growth}, we show that a $10^2 M_{\odot}$ stellar remnant formed at $z=30$ growing at an average Eddington ratio of 1.85 after a 100~Myr delay in starting accretion can easily reproduce the BH mass of \src{}. In reality, we expect that super-Eddington accretion occurs in short bursts \citep[e.g.,][]{takeo20,suh25}. However, we plot the simplified average Eddington ratio growth curve here for visual clarity.

\subsection{A Dense Gas Origin?}\label{sec:densegas}
\begin{figure*}[t]
\includegraphics[width=\textwidth]{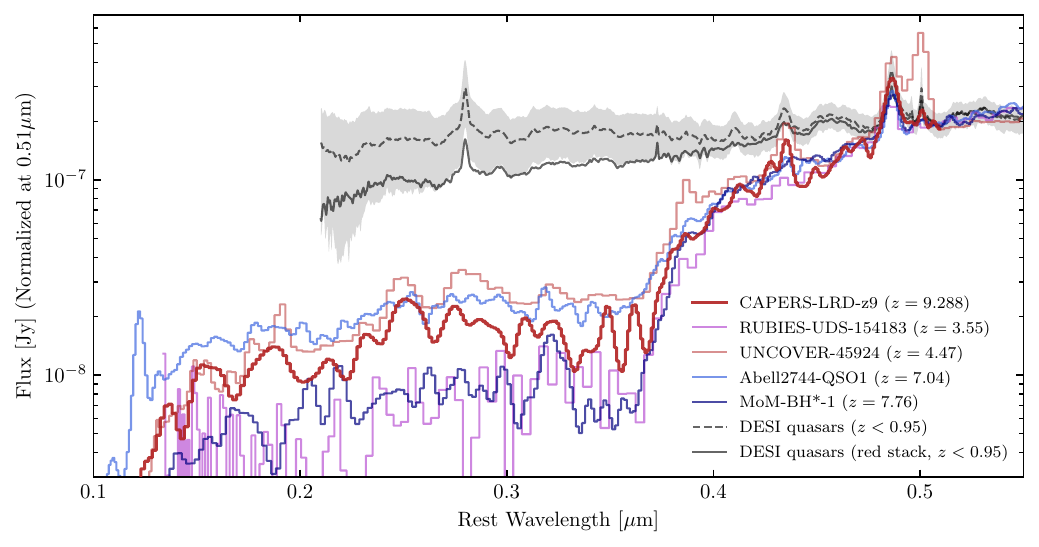}
\caption{The NIRSpec MSA/PRISM spectra of \src{}, MoM-BH*-1 at $z=7.76$ \citep{naidu25}, RUBIES-UDS-154183 at $z=3.55$ \citep{degraaff25}, UNCOVER-45924 at $z=4.47$ \citep{labbe24}, and Abell2744-QSO1 at $=7.04$ \citep{furtak24,ji25} normalized to the spectrum of \src{} at rest-frame $0.51\mu$m. We apply a $1$--$2$ pixel Gaussian smoothing (smoothing lower-$z$ objects less as they are already relatively smoothed by the PRISM's lower spectral resolution at bluer observed wavelengths) to all five spectra to increase visual clarity. For comparison, we plot the median and 10th--90th percentile of DESI spectra of 232 quasars ($z<0.95$) selected to have similar values of H$\beta_{broad}$ FWHM, $L$(\oiii$\lambda5007$), and \oiii$\lambda5007$/H$\beta$ flux ratio (gray shading and gray dashed curve), and the stack of the reddest 20 DESI quasars (gray solid curve). While the rest-optical emission from \src{} strongly resembles that of the other NIRSpec sources, these sources are all starkly different from quasars identified in DESI, where even the reddest DESI quasars fail to approach the reddening or Balmer break exhibited in the NIRSpec LRDs.}
\label{fig:break}
\end{figure*}

As referenced in \S\ref{sec:intro}, the ``dense-gas'' model of LRDs has garnered significant attention in the literature in recent months. This model, introduced by \cite{inayoshi25}, has recently been applied to two newly discovered LRDs---MoM-BH*-1 at $z=7.76$ \citep{naidu25} and RUBIES-UDS-154183 at $z=3.55$ \citep{degraaff25}---as well as the triply-imaged $z=7.04$ LRD Abell2744-QSO1 \citep{furtak24, ji25}. We plot these objects, and the $z=4.47$ LRD (UNCOVER-45924) from \cite{labbe24}, in Figure~\ref{fig:break}. All spectra are normalized to the spectrum of \src{} at rest-frame $0.51\mu$m.

All of these objects are notable for their strong Balmer breaks, which, in the cases of MoM-BH*-1 and RUBIES-UDS-154183, were insufficiently fit by evolved stellar populations \citep{naidu25,degraaff25}. The Balmer breaks and SEDs of both objects were better fit by invoking the dense-gas model of \citet{inayoshi25}. Similarly, while another object---A2744-45924---can be fit with a stellar model, the implied stellar mass is large ($\sim 10^{11}\,M_\odot$) and in tension with ALMA dynamical mass measurement $\sim 10^9\,M_\odot$ \citep{akins25b}. 

While \src{} shows a factor of $\sim1.5$ softer break $\left(f_{\nu,4050\textrm{\AA{}}}/f_{\nu,3670\textrm{\AA{}}}=4.35^{+0.93}_{-0.67}\right)$ than MoM-BH*-1 ($7.7^{+2.3}_{-1.4}$) and RUBIES-UDS-154183 ($6.9^{+2.8}_{-1.5}$), all of these objects show a remarkably similar rest-frame optical SED and Balmer break shape, suggesting common physical and emission structures. However, unlike MoM-BH*-1, \src{} demonstrates only moderate H$\beta$ absorption and shows a strong detection of H$\gamma$ in emission. Higher resolution spectroscopy of \src{} beyond that of NIRSpec/PRISM would allow for a higher fidelity examination of these features. 

Given the similarity in spectral shapes between \src{} and the sources in \cite{naidu25} and \cite{degraaff25}, we propose that the rest-frame optical light from \src{} is indeed dominated by AGN emission through a shell of dense ($\gtrsim10^{10}$~cm$^{-3}$) neutral gas, and that the rest-UV emission blue-ward of the Balmer break may be stellar emission from the host galaxy \citep[e.g.][]{kocevski23} or scattered light, originating from the AGN \citep[e.g.][]{greene24}. From our \texttt{Cloudy} modeling, we infer a gas density of $\log n_H/\textrm{cm}^{-3}=9.9^{+0.2}_{-0.2}$ and column density $\log N_H/{\rm cm}^{-2} \gtrsim 26$. These results are remarkably comparable to MoM-BH*-1 and RUBIES-UDS-154183, further suggesting that these objects may have similarities in their physical structure. However, we note that this dense regime is beyond \texttt{Cloudy}'s recommended use cases, and alternate analyses may be necessary in future work to fully understand the nature of these exotic sources.

We also note that our best-fit spectrophotometric model overpredicts the MIRI F1000W flux by $\chi = -1.8$. 
This data was used in the fitting procedure, but the model was not flexible enough to simultaneously fit the NIRSpec+MIRI data. We attempt to better match the MIRI photometry by including a flexible dust attenuation law for the AGN component of the model (see Appendix~\ref{app:dustlaw} for details). In this revised fit, we find a best fit \cite{salim18} $\delta$ parameter---which governs the steepness of the attenuation curve, with negative values producing increasingly steeper curves---of $\delta=-4.66^{+0.31}_{-0.24}$.  While the resulting fit matches the MIRI F1000W flux with $\chi=-0.27$ with no detrimental effects on the rest of the fit, $\delta=-4.66$ represents an attenuation curve that is unphysically steep, as it is by definition a \citep{calzetti01} curve multiplied by an additional factor of $(\lambda/5500\textrm{\AA})^{4.66}$. For reference, an SMC law can be approximated with a comparatively small $\delta=-0.45$. While steep attenuation curves were also explored in \cite{degraaff25} to reconcile their MIRI observations with the dense-gas LRD model, we agree with their assessment that such curves are likely unrealistic. Alternate models and modeling parameters are thus necessary to reproduce the rest-NIR properties of the LRDs, including---perhaps---modifying the intrinsic SED of the accretion disk. Such modeling may soon be robustly enabled by the increasing known population of dense-gas modeled LRDs.

In terms of the overall assembly history of the first supermassive black holes, we may witness a three-stage evolutionary sequence: the initial seeding process, a period of rapid growth during the LRD stage driven by massive inflows of gas, and a final `clearing of the fog' (possibly driven by radiation-hydrodynamics processes; e.g., \citealt{Smith_DCBH2017}), when the emerging SMBHs become unobscured, and with sufficient cosmic time to assemble additional stellar mass, join the `standard' quasar/AGN sequences at lower redshifts.
More specifically, \citet{inayoshi25b} propose that LRDs may be simply the first incidence of AGN activity in a galaxy's evolution, based on the log-normal occurrence rate of LRDs with cosmic time. 
The nascent evolutionary stage of the LRDs in this toy model may also explain their unique and extreme characteristics. 
The SMBH could form early in a galaxy's lifespan, surrounded by a large reservoir of dense neutral gas, which serves as both a readily available supply of accretion material and as the shell of dense gas surrounding the AGN (as in \citealt{inayoshi25}). 
This scenario would result in an over-massive SMBH producing a dense-gas reddened SED with a strong Balmer break, in combination with nascent star formation in the host galaxy that produces a blue rest-UV slope. 
In summation, these two components would supply the two halves of the defining LRD ``v-shaped'' SED for the duration of the initial ``LRD-phase'' of AGN activity. 
Given the low metallicity/dust content, but high column density of the nuclear gas, this model also favorably reproduces the lack of mid-infrared dust emission \citep{Williams24,PerezGonzalez24,akins24,leung24}, and lack of X-rays \citep{yue24xray,maiolino25, lambrides24} observed in LRDs. 
During future AGN active periods---after a period of dormancy---the stellar mass of the galaxy will have had sufficient time to accumulate and produce a more typical black hole to stellar mass ratio of $M_{\textrm{BH}}/M_{*}=0.1\%$. 
Simultaneously, the dense gas near the AGN would be consumed by the AGN during the LRD-phase, such that in future instances of AGN activity, the resulting evolved galaxy and AGN resemble a classical quasar/AGN. 

\subsection{A non-stellar origin for the UV continuum?}\label{sec:nonstellar}
In \S\ref{sec:stellarmass} we have estimated the stellar mass of the host galaxy of \src{} under the assumption that its UV luminosity is primarily powered by star formation. Even a visual inspection, however, shows that the shape of the UV continuum is unlike that of star-forming galaxies at comparable redshifts, \citep[e.g.,][]{castellano22,kokorev25}. As Figure~\ref{fig:break} illustrates, despite the relatively low SNR, the slope of the continuum of \src{} appears to change at around rest-frame $\approx 2500$~\AA{}, redward of which is relatively blue and rising toward shorter wavelength, whereas at bluer wavelengths it flattens and then perhaps decreases. 
Moreover, at rest-frame $\approx2500$\,\AA{} (observed-frame $2.6\,\mu$m) there is a hint of an apparent broad emission feature (see Figs.~\ref{fig:fullspectrum},\ref{fig:combinedfit}). 
Figure~\ref{fig:break} also shows an overall similarity between the UV continuum of \src{} and that of UNCOVER-45924, which has a substantially higher SNR, and strong UV lines such as C\,\textsc{iv}\,$\lambda 1550$ and C\,\textsc{iii}]\,$\lambda 1908$ and \Feii{} lines commonly seen in quasars \citep[e.g.][]{vestergaard01}.
These features are not detected with statistical significance in the spectrum of \src{}, yet the broad emission feature at $\approx2500$\,\AA\ may be associated with the \Feii{} pseudocontinuum, as in A2744-45924.
This, as well as the overall shape of the continuum, are intriguing and add credence to the idea that the UV continuum in \src{} could be nonstellar. 

To further explore this idea, we compare the LRDs to a control sample of quasars ($z<0.95$) from the Dark Energy Spectroscopic Instrument (DESI) DR1 dataset \citep{desi_dr1}. We assemble this sample by selecting DESI quasars with similar H$\beta_\textrm{broad}$ FWHM, $L$(\oiii$\lambda5007$), and \oiii$\lambda5007$/H$\beta_\textrm{total}$ flux ratios to \src{} based on \texttt{FastSpecFit} \citep{moustakas2023} emission-line measurements reported in the DESI AGN/QSO value-added catalog\footnote{\url{https://data.desi.lbl.gov/doc/releases/dr1/vac/agnqso/}} (S. Juneau et al., in preparation). We plot both the 10th--90th percentile of these objects as well as the inverse-variance weighted mean stack of the 20 reddest objects in Figure~\ref{fig:break} as a standard of comparison. It is immediately clear that even the stack of the reddest DESI quasars strongly differs from the LRD rest-optical continua, including a lack of the prominent \Feii{} bump at 4434–4684\AA{} in the LRDs (which is not expected to appear in faint [$L_{\textrm{H}\beta}<10^{44}$~erg~s$^{-1}$] high-redshift AGN e.g., \citealt{Trefoloni2024}). 
However, as mentioned above, in the rest-UV, \src{} and UNCOVER-45924 both exhibit possible AGN-signature emission features such as the $\sim0.25\mu$m iron bump, prominently featured in the DESI stack, suggesting that the rest-UV emission may indeed have an AGN origin.

The question naturally arises how UV emission from the accretion disk can escape such a dense gas cloud. As suggested by \citet{greene24}, the rest-UV light in LRDs could originate from accretion disk photons either being scattered or directly transmitted (via a patchy medium). If we simply re-scale the rest-UV continuum from the DESI quasars shown in Figure~\ref{fig:break} to match our data, we find that the UV slopes are roughly consistent with an implied scattering fraction of $\sim$10\%, fully consistent with the results presented in \citet{greene24}. The scattering medium could be neutral (Rayleigh scattering) or partially ionized gas (Rayleigh and Thomson scattering), and could be associated with either the external regions of the same dense gaseous envelope that enshrouds the AGN or a different regions altogether. The type of scattering medium and its geometry will, in general, contribute to the overall spectral shape and intensity, given the specific wavelength-dependence of the scattered radiation, and this might also help explain the diversity of spectral morphologies observed this category of sources. 

If the UV emission is predominantly of AGN origin, the upper limit to the stellar mass discussed in \S\ref{sec:stellarmass} will have to be revised downward by a substantial amount, which is unconstrained by the current data.  This makes the ratio $M_{BH}/M_*\gg 4.5\%$ substantially more extreme, with important implications on the mechanisms of formation and co-evolution of SMBHs and their host galaxies.

\section{Summary}\label{sec:summary}

In this work we spectroscopically confirm and analyze the highest redshift BLAGN observed to date: \src{}. We summarize our primary results below:
\begin{enumerate}
\item{Using \textit{JWST}/NIRSpec PRISM spectroscopy from the CAPERS program, we identify the highest redshift ($z=9.288$) BLAGN yet discovered---\src{}---based on a strong detection of broad (FWHM=$3525\pm589$~km~s$^{-1}$) H$\beta$ emission.}
\item{We determine that \src{} is unresolved in NIRCam imaging, placing upper limits of $\lesssim 175$ pc and $\lesssim 350$ pc on its physical size in the rest-UV and rest-optical.}
\item{We use the \citet{greene05} empirical relations to estimate a black hole mass of $\log\left(M_{\textrm{BH}}/M_{\odot}\right)=7.58\pm0.15$. We place conservative bounds on this mass measurement, accounting for systematic uncertainties, by assuming Eddington-limited accretion (lower bound) and applying a dust correction based on the Balmer decrement (upper bound). This yields a range of $\log\left(M_{\textrm{BH}}/M_{\odot}\right)=6.65$--$8.50$.}
\item{We successfully model \src{} as an AGN enshrouded in a shell of dense ($n_H > 10^{9}$~cm$^{-3}$) neutral gas using \texttt{Cloudy} following the models of \citet{inayoshi25} that have been recently applied to similar sources \citep{ji25,naidu25,degraaff25}. We fit the rest-frame UV to a stellar population model using \texttt{bagpipes}. From this modeling, we derive an upper limit on the host galaxy stellar mass of $<10^9M_{\odot}$ and estimate a neutral gas density near the AGN of $n_H \sim 10^{10}$ cm$^{-3}$.}
\item{From our canonical black hole mass, we calculate a black hole mass to stellar mass ratio $M_{\textrm{BH}}/M_*>4.5\%$ for \src{}. However, we recognize our estimates of both $M_\textrm{BH}$ and $M_*$ are subject to large systematic uncertainties.  Based on our systematic bounds of $M_{\textrm{BH}}$, the lower limit on $M_{\textrm{BH}}/M_*$ ranges from $>$46\% to as low as $>$0.5\%. Even our absolute lowest bound presents a significant deviation from the ``typical'' $M_\textrm{BH}/M_*\sim0.1\%$ value exhibited by local quasars.}
\item{We model the growth of the SMBH powering \src{} and find that an Eddington-limited heavy ($\sim$\,10$^{4}M_{\odot})$ seed or a super-Eddington light ($\sim$\,10$^{2}M_{\odot})$ seed are necessary to produce such a SMBH by $z=9.288$.}
\end{enumerate}

From these results, it is clear that \src{} is an extreme example of an LRD in redshift, $M_\textrm{BH}/M_*$, and neutral gas density. Along with MoM-BH*-1 and RUBIES-UDS-154183, \src{} provides strong evidence in support of the ``dense-gas enshrouded AGN'' physical explanation for the rest-frame optical emission from LRDs. Furthermore, its unprecedented redshift for a BLAGN provides insight into the interplay between black hole formation and growth and the physical properties of LRDs. While we present an initial analysis of \src{} in this work, further analysis of such an exotic source will undoubtedly continue to enhance our understanding of galaxy, AGN, and SMBH evolution in the very early universe. 

\begin{acknowledgements}
AJT acknowledges support from the UT Austin College of Natural Sciences, and AJT and SLF acknowledge support from STScI/NASA through JWST-GO-6368. ACC and HL acknowledge support from a UKRI Frontier Research Guarantee Grant (PI Carnall; grant reference EP/Y037065/1). LN acknowledges support from grant ``Progetti per Avvio alla Ricerca - Tipo 1, Unveiling Cosmic Dawn: Galaxy Evolution with CAPERS" (AR1241906F947685). JSD acknowledges the support of the Royal Society via the award of a Royal Society Research Professorship. RA acknowledges support of Grant PID2023-147386NB-I00 funded by MICIU/AEI/10.13039/501100011033 and by ERDF/EU, and the Severo Ochoa grant CEX2021-001131-S funded by MCIN/AEI/10.13039/50110001103. FC and TMS acknowledge support from a UKRI Frontier Research Guarantee Grant (PI Cullen; grant reference EP/X021025/1). K.I. acknowledges support from the National Natural Science Foundation of China (12233001),  the National Key R\&D Program of China (2022YFF0503401), and the China Manned Space Program (CMS-CSST-2025-A09).

This work is based on observations made with the NASA/ESA/CSA \textit{James Webb Space Telescope}, obtained at the Space Telescope Science Institute, which is operated by the Association of Universities for Research in Astronomy, Incorporated, under NASA contract NAS5-03127. Support for program number GO-6368 was provided through a grant from the STScI under NASA contract NAS5-03127. The data were obtained from the Mikulski Archive for Space Telescopes (MAST) at the Space Telescope Science Institute. 
These observations are associated with program \#6368, and can be accessed via \dataset[doi: 10.17909/0q3p-sp24]{http://dx.doi.org/10.17909/0q3p-sp24}.

This research used data obtained with the Dark Energy Spectroscopic Instrument (DESI). DESI construction and operations is managed by the Lawrence Berkeley National Laboratory. This material is based upon work supported by the U.S. Department of Energy, Office of Science, Office of High-Energy Physics, under Contract No. DE–AC02–05CH11231, and by the National Energy Research Scientific Computing Center, a DOE Office of Science User Facility under the same contract. Additional support for DESI was provided by the U.S. National Science Foundation (NSF), Division of Astronomical Sciences under Contract No. AST-0950945 to the NSF National Optical-Infrared Astronomy Research Laboratory; the Science and Technology Facilities Council of the United Kingdom; the Gordon and Betty Moore Foundation; the Heising-Simons Foundation; the French Alternative Energies and Atomic Energy Commission (CEA); the National Council of Humanities, Science and Technology of Mexico (CONAHCYT); the Ministry of Science and Innovation of Spain (MICINN), and by the DESI Member Institutions: www.desi.lbl.gov/collaborating-institutions. The DESI collaboration is honored to be permitted to conduct scientific research on I’oligam Du’ag (Kitt Peak), a mountain with particular significance to the Tohono O’odham Nation.

MED thanks Arjun Dey for helpful early comments about the spectrum of \src{}. We also acknowledge Haojie Hu for kindly sharing the data from their black hole seeding and growth model, which we use in Figure~\ref{fig:mbhmstar} for comparison with our observational results.
\end{acknowledgements}


\facilities{JWST}

\software{astropy: \cite{astropy:2013,astropy:2018,astropy22}, bagpipes: \cite{carnall18}, Cloudy: \cite{cloudy23}, emcee: \cite{emcee}, JWST pipeline: \cite{bushouse23}, scipy: \cite{scipy20}, SPARCL: \cite{juneau2024}}

\bibliographystyle{aasjournalv7.bst}
\bibliography{ref1.bib}

\appendix
\section{Flexible Dust Law Fit}\label{app:dustlaw}
\begin{figure*}[ht]
\includegraphics[width=\textwidth]{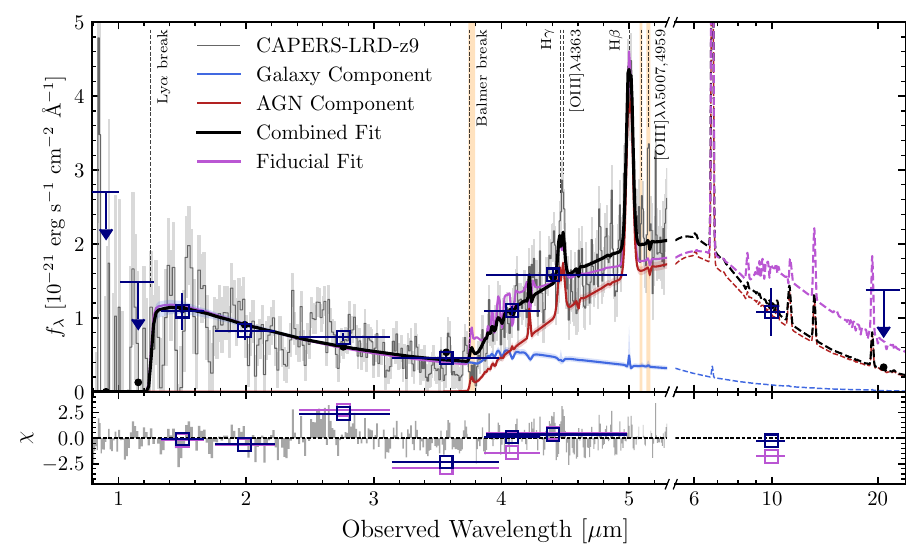}
\caption{Spectrum and 1$\sigma$ errors of \src{} (gray curve, light gray shading), the best fit host galaxy component (blue and dashed blue curves), the best fit dense-gas enshrouded AGN component (red and dashed red curves), combined host+AGN fit (black and dashed black curves), fidicuial SMC dust law fit (purple and dashed purple curves), photometry data (dark blue squares and upper limits), best-fit model photometry (black points) in the upper panel, and masked regions (gold shading). The lower panel shows the $\chi$ residuals of the fit for the spectrum (gray curve) and photometry (dark blue squares). Note that the flexible (yet unphysically steep) AGN dust law allows for strong agreement with the MIRI F1000W flux that cannot be reproduced by an SMC law.}
\label{fig:combinedfitflexibledust}
\end{figure*}

Here we describe our fitting procedure and results for a flexible \citep{salim18} dust law. We adopt all of the same parameters and procedures described in \S\ref{sec:templates} with the exception of the AGN component dust here. We substitute the fixed SMC dust law for a flexible slope \citep{salim18} dust law. We add the steepness parameter $\delta$ to the model parameters and assume a flat prior ranging from $\delta=-5$ to $0$. We show the results of this updated fitted model in Figure~\ref{fig:combinedfitflexibledust} and Table~\ref{tab:modelsflexibledust}.

The resulting model better reproduces the MIRI F1000W photometry (with residuals $\chi=-0.27$ versus $\chi=-1.8$ with the SMC law). However the best-fit $\delta=-4.66^{+0.31}_{-0.24}$ represents an unphysically steep attenuation curve (see discussion in \S\ref{sec:densegas}). Clearly more sophisticated modeling is necessary to reproduce the SEDs of LRDs--including perhaps, modifications the the intrinsic SED of the AGN accretion disk.

\begin{deluxetable}{@{\extracolsep{10pt}}l@{}c@{}c@{}}
\caption{\texttt{Cloudy} parameter grid \& fitted values}\label{tab:modelsflexibledust}
\tablehead{Parameter & Grid Values & Fitted Value}
\startdata
$\log T_{\rm BB}$/K & $4$, $4.7$, $5$, $5.7$ & $5.0$ \cr
$\alpha_{\rm OX}$ & $-2.5$, $-2.0$, $-1.5$ & $-1.5$ \cr
$\alpha_{\rm UV}$ & $-0.1$ & \dots \cr
$\alpha_{\rm X}$ & $-0.5$ & \dots \cr
$\log n_H/{\rm cm}^{-3}$ & $9, 9.5, 10, 10.5, 11, 11.5, 12$ & $10.3^{+0.3}_{-0.3}$ \cr 
$v_{\rm turb}/{\rm km}\,{\rm s}^{-1}$ & $100, 200, 300, 400, 500$ & $134^{+45}_{-26}$ \cr
[Fe/H] & $-2$ & \dots \cr
$\log U$ & $-3.5, -3, -2.5, -1.5, -0.5$ & $-1.5$
\cr
$\log N_H/{\rm cm}^{-2}$ & $21, 22, 23, 24, 25, 26$ & $25.0^{+0.3}_{-0.5}$ \cr
\hline 
$C_f$ & 0--1 & $0.64^{+0.14}_{-0.13}$ \cr 
$A_V$ & 0--3 (\citealt{salim18} law) & $0.32^{+0.05}_{-0.04}$ \cr
$\delta$ & $-5$--$0$ & $-4.66^{+0.31}_{-0.24}$
\enddata
\tablecomments{The first four parameters describe the incident continuum spectrum where $T_{\rm BB}$ is the `Big Bump' temperature, $\alpha_{\rm OX}$ is the optical to X-ray index, $\alpha_{\rm UV}$ is the power-law UV slope and $\alpha_{\rm X}$ is the power-law X-ray slope. The incident continuum is passed through dense gas defined by the gas density ($n_H$), metallicity ([Fe/H]) and turbulent velocity ($v_{\rm turb}$). The level of irradiation at the face of the cloud is set by the ionization parameter ($\log U$) and the \texttt{Cloudy} calculations are stopped at a range of line-of-sight column densities ($N_H$). $\delta$ is the \cite{salim18} dust law steepness parameter.}
\end{deluxetable}

\end{document}